\documentclass[a4paper,fleqn]{cas-sc}
\usepackage{threeparttable}
\usepackage[ruled,linesnumbered]{algorithm2e}
\usepackage{amsmath}
\usepackage{amssymb}

\usepackage[authoryear]{natbib}

\def\tsc#1{\csdef{#1}{\textsc{\lowercase{#1}}\xspace}}
\tsc{WGM}
\tsc{QE}

\begin{document}
\let\WriteBookmarks\relax
\def\floatpagepagefraction{1}
\def\textpagefraction{.001}

\shorttitle{}    

\shortauthors{}  

\title [mode = title]{A decomposition-integration framework coupling singular spectrum analysis with a volatility-adaptive LSTM for stock price forecasting}

\tnotemark[1] 

\tnotetext[1]{} 

%

\author{Xin-Rui Lu}

\credit{Methodology, Investigation, Software, Data curation, Validation, Writing - original draft}

\affiliation{organization={School of Science},
            addressline={University of Science and Technology Liaoning}, 
            city={Anshan},
            postcode={114051}, 
            state={Liaoning},
            country={PR China}}

\author{Da-Qing Zhang}[
  orcid=0000-0001-6648-5360     
]
\cormark[1]

\ead{d.q.zhang@ustl.edu.cn}


\credit{Conceptualization, Methodology, Investigation, Validation, Writing - review \& editing}


\cortext[1]{Corresponding author}

\fntext[1]{}


\begin{abstract}
Financial time series exhibit multi-scale, high-noise, and non-stationary volatility dynamics. Decomposition--integration frameworks built on complete ensemble empirical mode decomposition with adaptive noise (CEEMDAN) have advanced this field, yet existing CEEMDAN hybrids remain unstable: high-frequency mode mixing contaminates downstream predictors, and the heterogeneous volatility of decomposed components is not matched by any volatility-adaptive mechanism. To address these gaps, this paper proposes CEEMDAN-SE-SSA-VMGL. CEEMDAN adaptively decomposes the original series, and sample-entropy-driven K-means clustering groups the intrinsic mode functions into high-, mid-, and low-frequency and residual components. Singular spectrum analysis (SSA) is then applied to the high-frequency group, whose singular-value decomposition of the trajectory matrix and principal-component reconstruction decouple intertwined pseudo-oscillatory modes and alleviate mode mixing. We further design a plug-and-play robust LSTM variant, the volatility-modulated GARCH-LSTM (VMGL), which embeds a differentiable GARCH(1,1) layer inside the LSTM cell as a risk gate; the estimated conditional volatility multiplicatively scales the forget-gate activation, letting memory decay adapt to the instantaneous volatility while remaining fully end-to-end trainable. Component predictions are linearly combined. On five US and Chinese financial series (CSI 300, CSI 500, AAPL, GOOGL, S\&P 500), the proposed model is compared with eleven benchmark architectures, lowering the forecasting error by 9\% to 24\% relative to the best benchmark and by 60\% to 70\% relative to a standard LSTM. Diebold--Mariano and model-confidence-set (MCS) tests confirm the statistical significance and stability of the gains, spectral analyses verify the SSA denoising efficacy, and backtesting demonstrates the framework's economic utility.
\end{abstract}


\begin{highlights}
\item Component-wise volatility-adaptive modeling: a new decomposition-integration paradigm.
\item Applying SSA to high-frequency financial components alleviates mode mixing.
\item A plug-and-play volatility-adaptive robust LSTM variant improves forecast stability.	
\item Achieves the best performance among eleven benchmark models on all five datasets.
\item Diebold--Mariano tests, MCS and backtesting confirm statistical and economic value.
\end{highlights}

\begin{keywords}
	Financial time series forecasting \sep Hybrid model \sep Complete ensemble empirical mode decomposition \sep Singular spectrum analysis \sep GARCH \sep Long short-term memory 
\end{keywords}

\maketitle

\section{Introduction}
\label{sec1}
Financial time-series forecasting is a central task in modern finance, with direct implications for portfolio management, systemic-risk control, and macroprudential policy design \citep{Vancsura2025,Alzaman2024,Jang2023}. Stock prices and market indices exhibit multi-scale dynamics, high-frequency noise, non-stationary volatility clustering, and nonlinear dependence, and these properties fundamentally resist characterization under linear statistical assumptions \citep{Zhao2024,He2016,Wei2025}. Such complexities stem from the heterogeneous behavior of market participants and from the continuous information flow among global exchanges, and they demand forecasting frameworks able to disentangle multi-resolution signal components while adapting to rapidly evolving market regimes \citep{Mutinda2026,Gao2025,Yu2024}. Accurate and robust financial-market forecasting has therefore remained an active research topic across econometrics, statistical signal processing, and machine learning \citep{Ali2025,Jiang2021,Yongchareon2025}.

Econometric approaches played a key role in early financial time-series forecasting. Among the most widely used are ARIMA and GARCH-type models, which provide standard tools for modeling linear autocorrelation and conditional heteroskedasticity, respectively. \citet{Dong2020} validated the predictive capability of ARIMA through extensive rolling-window experiments, while \citet{EKINCI2021} and \citet{Hassan2026} confirmed the effectiveness of GARCH variants in capturing asymmetric volatility effects under different market conditions. Nevertheless, these models impose strict stationarity assumptions and possess limited capacity for long-range nonlinear dependence, and such limitations tend to erode their reliability during structural breaks and market crises \citep{Alfeus2024,Maaz2023,Helena2022}.

Machine-learning models subsequently emerged as flexible nonparametric alternatives. \citet{Lin2025} combined CatBoost with metaheuristic optimization for multi-market price prediction; \citet{Wang2025} improved random forest through particle-swarm feature selection; and \citet{Benyan2023} built an interpretable financial-stability early-warning system using XGBoost with SHapley Additive exPlanations (SHAP) analysis. Despite these advances, such techniques generally rely on handcrafted feature engineering and exhibit fragile generalization under extreme volatility or abrupt distributional shift \citep{Wang2024,Xinzhu2023}.

Deep learning has become a powerful paradigm for end-to-end temporal representation learning. \citet{Hochreiter1997} introduced LSTM networks to mitigate gradient instability, and \citet{Wen2024} demonstrated that LSTM outperforms CNN and ANN in capturing long-term dependencies. \citet{Tashreef2023} integrated Transformer models with time2vec encoding, \citet{Ali2025} designed a cross-market graph neural network to exploit spatial dependencies among assets, and \citet{Yongchareon2025} carried out comprehensive benchmarking and showed that advanced architectures such as PatchTST perform well on short-horizon forecasting. However, when such models are deployed directly on raw price series, they fail to distinguish high-frequency noise from genuine market signals and memorize both indiscriminately. This behavior leads to overfitting and unstable prediction trajectories, especially under regime-switching volatility \citep{Wei2017,Gong2024}.

To address these limitations, hybrid decomposition--integration frameworks have become a widely adopted forecasting paradigm. These methods first decompose the raw series into simpler subcomponents through empirical mode decomposition or its variants, thereby reducing the complexity of the original signal and allowing each subcomponent to be modeled by a dedicated predictor. \citet{Torres2011} introduced CEEMDAN, which alleviates mode mixing through stagewise adaptive noise injection, and subsequent studies coupled this decomposition front-end with diverse forecasting architectures. \citet{Mutinda2026} proposed the CEEMDAN-SE-VMD-BiGRU-Ridge framework, \citet{Gao2025} developed a CEEMDAN-Informer model for long-horizon forecasting of Chinese A-shares, and \citet{Shen2026} designed a multi-channel complementary network with Transformer residual prediction. \citet{Li2025} further differentiated predictor selection according to the frequency characteristics of each component, while \citet{Ma2025} combined wavelet decomposition with a multi-task self-attention mechanism. Although such frameworks have consistently improved the forecasting accuracy of financial time series over the past decade, their stability across different volatility regimes has not improved correspondingly. Growing evidence indicates that when markets shift from calm to high-volatility conditions, even carefully tuned hybrid models may generate strongly oscillating prediction trajectories and, in some cases, become less stable than a single recurrent network applied without decomposition \citep{Zhu2024,Yu2025,Ouyang2026}.

These limitations call for targeted interventions at both the data and the model levels. On the data side, additional denoising should be applied selectively to the high-frequency components most seriously contaminated by mode mixing, rather than filtering all components indiscriminately. On the model side, the predictor should perceive time-varying volatility and adjust its memory behavior according to the current market state. Following this rationale, this paper proposes CEEMDAN-SE-SSA-VMGL, a hybrid framework that couples targeted data-side signal purification with model-side volatility-adaptive memory adjustment within a unified decomposition--integration pipeline. The framework proceeds in three steps. First, CEEMDAN adaptively decomposes the original series into intrinsic mode functions, which are aggregated by sample-entropy-based K-means clustering into four component groups: the high-, mid-, and low-frequency components, together with the residual. Second, singular spectrum analysis is applied to the high-frequency group, performing a singular-value decomposition of the trajectory matrix to decouple the tangled pseudo-oscillatory modes and to suppress the noise remaining after the main decomposition. Third, we introduce the volatility-modulated GARCH-LSTM (VMGL) for the dominant oscillatory components, embedding a differentiable GARCH(1,1) layer inside the recurrent unit as an internal risk gate. This design enables end-to-end joint optimization of conditional-variance estimation and sequence prediction. The estimated volatility then dynamically scales the forget-gate activation, allowing the model to accelerate memory decay in turbulent periods and to retain long-term dependence in calm regimes. The final forecast is obtained by linearly summing the component predictions. The proposed framework is evaluated on five representative financial series spanning the US and Chinese markets and is compared against eleven benchmark architectures.

The contributions of this study are fourfold.
\begin{enumerate}
\item \textbf{A novel data-side signal-purification strategy.} A hybrid signal-processing pipeline integrating CEEMDAN, sample-entropy-based clustering, and singular spectrum analysis is proposed. In this pipeline, SSA is applied specifically to the high-frequency group, and a singular-value decomposition of its trajectory matrix decouples the tangled pseudo-oscillatory modes. Spectral analysis confirms that this targeted denoising effectively suppresses mode mixing and yields purer components for downstream modeling.
\item \textbf{A robust LSTM variant with a learnable GARCH risk gate.} VMGL, a volatility-modulated GARCH-LSTM, is presented, embedding a differentiable GARCH(1,1) layer inside the LSTM cell as an internal risk gate. The GARCH coefficients $\omega$, $\alpha$, and $\beta$ are learnable and are optimized by backpropagation together with the recurrent weights, enabling joint end-to-end training of conditional-variance estimation and sequence prediction. The estimated volatility then multiplicatively scales the forget-gate activation, letting memory retention adapt to the prevailing market regime.
\item \textbf{Comprehensive multi-metric evaluation and benchmarking.} The framework is evaluated on five major financial series spanning the US and Chinese markets, including CSI 300, CSI 500, AAPL, GOOGL, and S\&P 500. Its performance is benchmarked against eleven competing models, including standalone recurrent networks, single-decomposition, and secondary-decomposition hybrid frameworks. The proposed model reduces the forecasting error by 9\% to 24\% relative to the best-performing competitor and by 60\% to 70\% relative to a standard recurrent benchmark.
\item \textbf{Rigorous statistical and economic validation.} Beyond standard error metrics, the statistical superiority of the proposed model is verified through Diebold--Mariano and model-confidence-set tests. Power-spectral-density and Hilbert marginal spectrum analyses jointly corroborate the denoising efficacy of the signal-purification module. Portfolio backtesting with cumulative-return analysis demonstrates the practical reliability and economic utility of the proposed framework in volatile financial environments.
\end{enumerate}

The remainder of this paper is organized as follows. Section~\ref{sec2} reviews the literature on financial time-series forecasting. Section~\ref{sec3} details the architecture of the proposed CEEMDAN-SE-SSA-VMGL framework. Section~\ref{sec4} describes the experimental setup on five US and Chinese financial series and presents the main empirical results. Section~\ref{sec5} summarizes the key findings, limitations, and directions for future work.

\section{Literature review}
\label{sec2}
Financial time-series forecasting has developed a broad family of models, each responding to the nonlinearity, non-stationarity, and volatility of financial series. These models can be broadly classified into four categories: traditional statistical models, machine-learning models, deep-learning techniques, and hybrid-model approaches. Although each category has its own strengths and limitations, hybrid frameworks have recently gained popularity because they integrate the complementary strengths of different methods and achieve superior predictive accuracy and robustness. This section briefly reviews each category, with emphasis on the hybrid approaches most relevant to the present study.

\subsection{Econometric models for financial market forecasting}
Early research on financial time-series forecasting relied mainly on econometric and statistical models, with ARMA, ARIMA, and GARCH among the commonly used techniques \citep{EKINCI2021}. \citet{Hassan2026} proposed the NF-GARCH framework, which couples normalizing flows with GARCH to learn the residual distribution flexibly; the framework markedly improves the predictive accuracy of conventional GARCH models on financial time series. \citet{Alfeus2024} compared HAR, realized GARCH, RECH, and RFSV models and found that realized GARCH effectively captures volatility clustering and persistence, offering important guidance for risk management in emerging markets. ARIMA, another classical time-series approach, has also been widely applied to financial price forecasting. \citet{Liu2022} used ARIMA to forecast the unit price and price changes of Bitcoin and gold, achieving satisfactory results. \citet{Dong2020} evaluated the predictive ability of ARIMA on daily US stock returns through more than two million rolling-window forecasts, finding that a five-year fixed-width window fits best and that the lack of nonlinearity in ARIMA was not its main drawback. Although traditional statistical models show some practical value, their excessive reliance on stationarity and their difficulty in capturing long-range dependence limit their performance on complex time-series-analysis tasks.

\subsection{Machine learning models for financial market forecasting}
To improve predictive accuracy, researchers have applied various machine-learning techniques across multiple markets and datasets. Some studies focus on enhancing model performance through parameter optimization and ensemble learning. \citet{Lin2025} combined CatBoost with the Marine Predators Algorithm for Google stock-price forecasting, obtaining $R^2$ values above 0.99 on all four markets (DAX, FTSE, HSI, and SSE). \citet{Wang2025}, addressing the parameter-optimization deficiencies and feature-selection difficulties of random forest in stock rise/fall prediction, proposed an improved random-forest algorithm combining particle-swarm feature selection with grid-search parameter optimization; the resulting AUC exceeds those of the original random forest, decision tree, and support vector machine. Other studies focus on comparative analysis. \citet{Tas2022} compared nonlinear autoregressive (NARX) networks with SVR on high-trading-volume stock-price prediction at the Istanbul Stock Exchange, finding both to be strongly predictive yet each with distinct advantages. \citet{Benyan2023} constructed a financial-stability index (ICFS) from 20 daily indicators across the Chinese money, stock, bond, and foreign-exchange markets and used an XGBoost model for financial-risk early warning, showing that XGBoost outperforms other machine-learning models. Although machine-learning models achieve notable accuracy gains over traditional statistical methods, their reliance on feature engineering, limited interpretability, and constrained adaptation to extreme volatility and structural breaks continue to restrict their wider application to complex financial time-series forecasting \citep{Vancsura2025}.

\subsection{Deep learning models for financial market forecasting}
Deep learning has been widely applied to stock-market prediction because of its ability to model the temporal dependencies and nonlinear dynamics of financial time series effectively. Early deep-learning research centered on recurrent neural networks and their variants. \citet{Lu2024} used a simple RNN and an LSTM to forecast the CSI 300 and NASDAQ indices, finding that both perform better on the CSI 300. \citet{Wen2024} compared CNN, LSTM, and ANN for stock-price prediction and found LSTM superior in capturing long-term dependence. To further optimize LSTM, \citet{Liu2025b} built a complex network from the net inflow of main capital in the stock market and extracted features such as node degree and betweenness centrality as additional inputs, markedly reducing prediction error. As research progressed, bidirectional recurrent structures and temporal convolution networks were introduced to better exploit temporal information. \citet{Mo2022} applied a bidirectional LSTM (BiLSTM) to financial time-series forecasting and found it more accurate than unidirectional LSTM, SVR, and ARIMA. The emergence of attention mechanisms and Transformer architectures then provided a new paradigm for time-series modeling. \citet{Tashreef2023} first integrated a Transformer with time2vec time encoding for stock-price prediction, outperforming ARIMA. Graph neural networks have also attracted attention for capturing spatial dependencies among assets. \citet{Ali2025} proposed the CMGM model, which designs hypergraphs and subgraphs to capture intra- and cross-market stock-price dependencies, achieving the lowest MAE (0.01148) and MSE (0.00026) over multiple markets. \citet{Cao2026} proposed the TimeKAN architecture, which captures complex temporal patterns through adaptive frequency decomposition and multi-order Kolmogorov--Arnold networks, achieving an average RMSE improvement of 21.5\% across several stock datasets. However, a single deep-learning model remains prone to overfitting and may fail to adapt quickly to changing market conditions \citep{Gong2024}. This limitation has shifted research attention toward hybrid and decomposition-based architectures.

\subsection{Hybrid models for financial market forecasting}
Hybrid forecasting has become an important paradigm in financial time-series forecasting. By complementing the strengths of multiple models, hybrid approaches construct more powerful predictors and achieve higher accuracy and robustness when handling stock-market data characterized by non-stationarity and volatility.

Hybrids that combine traditional econometric models with deep-learning models capture both linear and nonlinear features in the data. \citet{Zou2025} constructed an ARIMA-LSTM hybrid model to forecast Apple's stock price and found it markedly superior to the individual models in both RMSE and directional accuracy. \citet{Eskandari2026} proposed an ARIMA-GARCH-LSTM hybrid framework for value-at-risk (VaR) forecasting of WTI crude-oil returns, with the ARIMA-eGARCH-LSTM model achieving the lowest loss rate across all tests. \citet{Chen2025} further introduced sentiment analysis, using BERT to generate sentiment factors and combining GARCH parameters with an LSTM into an SF-GARCH-LSTM model that significantly improved predictive accuracy. \citet{Mei2026} built a Chinese stock-market uncertainty index via LASSO and PCA, embedded GRU, LSTM, and BiLSTM into a GARCH-MIDAS model, and showed empirically that the ensemble is significantly more predictive than traditional models.

Integrating multiple machine-learning and deep-learning models enhances predictive robustness. \citet{Pham2022} designed an XGBoost-LSTM hybrid forecasting framework that extracts key features with XGBoost and captures temporal dynamics with LSTM, markedly outperforming the traditional ARIMA on foreign-exchange datasets. \citet{Alarbi2025} designed the AR-ERLM model, fusing ARIMA, RNN, and LightGBM with technical indicators and Google Trends data, outperforming several baselines on Netflix, Amazon, and Meta datasets. More recently, the rise of attention mechanisms and multi-scale architectures has encouraged attempts to combine Transformers and temporal convolution networks with recurrent networks. \citet{Tian2025} proposed the BiMT-TCN model, combining a bidirectional LSTM, an improved Transformer, and a temporal convolution network (TCN), achieving $R^2$ of 0.9779, 0.9776, and 0.9969 on the Shanghai Composite, Hang Seng, and NASDAQ indices, respectively. \citet{Chen2024} proposed MKP-TemporalNet, which dynamically calibrates the importance of multivariate time-series features through attention and improves TCN with a multi-kernel parallel convolutional structure combined with bidirectional GRU, validating its effectiveness on six new-energy stock indices.

Decomposition-based hybrid models have attracted growing attention in financial forecasting because of their unique advantage in handling data characterized by non-stationarity and volatility. Such methods generally first decompose the raw series into several intrinsic mode components or subsequences at the data-preprocessing stage, using algorithms such as empirical mode decomposition, wavelet transform, or variational mode decomposition; they then apply a predictor suited to each component's own characteristics and finally integrate the component predictions into the final output \citep{Shen2026,Mutinda2026}. This decomposition strategy reduces the complexity of the raw financial series, highlights local fluctuation regularities, and thereby improves overall predictive accuracy and stability.

Several studies employ single-layer empirical mode decomposition and its improved versions together with traditional machine learning or recurrent neural networks. \citet{Mutinda2026} proposed the CEEMDAN-SE-VMD-BiGRU-Ridge framework, which first decomposes the series with CEEMDAN, groups the modes into high-, mid-, and low-frequency components plus a residual through sample entropy and K-means clustering, further decomposes the high-frequency group with VMD, and models the resulting groups with BiGRU and ridge regression, achieving $R^2$ above 0.994 on six major global stock indices. \citet{Gao2025} proposed the StockCI model, coupling CEEMDAN with an Informer architecture: CEEMDAN adaptively decomposes the raw series, and Informer's ProbSparse self-attention performs long-horizon prediction for each component, reducing MAE by more than 15\% on average relative to the standalone Informer on high-frequency Chinese A-share data. \citet{Ouyang2026} fused EEMD, GRU, and Informer for S\&P 500 return forecasting, reducing RMSE by 61.4\% and MAE by 61\% relative to a traditional recurrent architecture. \citet{Yu2025} designed the EEMD-SE-GA-GRU framework, combining EEMD decomposition, sample-entropy clustering, and genetic-algorithm optimization of GRU hyperparameters, achieving $R^2=0.9814$ and MAPE=0.55\% on the Dow Jones Industrial Average. \citet{Zhao2025} proposed the ETT model, which uses CEEMD to decompose stock data, combines Time2Vec to capture periodic and aperiodic patterns, and learns long-range dependence through a Transformer, reducing MSE by 4\% on average and boosting cumulative returns by 58\% on the CSI 100 and CSI 300 datasets. \citet{Zhu2024} constructed the CEEMDAN-S-C-LSTM model, which introduces Savitzky--Golay filtering for denoising on the basis of CEEMDAN decomposition and combines a three-layer convolution with an LSTM network, reducing MAE by 45.33\% and RMSE by 43.44\% on the CSI 300 index relative to the best benchmark. Other studies focus on variational mode decomposition and its optimization, coupled with convolutional structures or TCN. \citet{Zhang2025} proposed VMD-ConvTSMixer, which decomposes signals into stationary components with VMD and captures temporal patterns and cross-channel dependencies with a convolution-enhanced TSMixer, reducing RMSE by 15\%--36\% relative to FEDformer and CrossFormer across several domain datasets. \citet{Yu2025b} proposed the GA-VMD-TCN method with genetic-algorithm-optimized VMD parameters for multi-step forecasting of the SSEC and S\&P 500 indices, consistently outperforming comparison models under alternative sample periods, evaluation methods, and business cycles.

Although decomposition--integration hybrid models have recently dominated financial time-series forecasting because of their effectiveness in handling non-stationary data, a critical gap remains in the literature. Existing methods generally omit dedicated denoising of the high-frequency intrinsic mode functions generated by CEEMDAN, so the pseudo-oscillations introduced by mode mixing are passed directly to the downstream predictors without suppression. At the same time, these predictors lack an adaptive mechanism for time-varying volatility, and residual noise is therefore amplified through recurrent iterations, leading to severe trajectory oscillation and predictive instability that remain unresolved. To overcome this limitation, this paper proposes the CEEMDAN-SE-SSA-VMGL framework, which couples SSA-based purification of the high-frequency group with VMGL, the volatility-modulated GARCH-LSTM, to stabilize forecasting across market regimes.

\section{Methodology}
\label{sec3}
\begin{figure}
	\centering
	\includegraphics[width=0.85\textwidth]{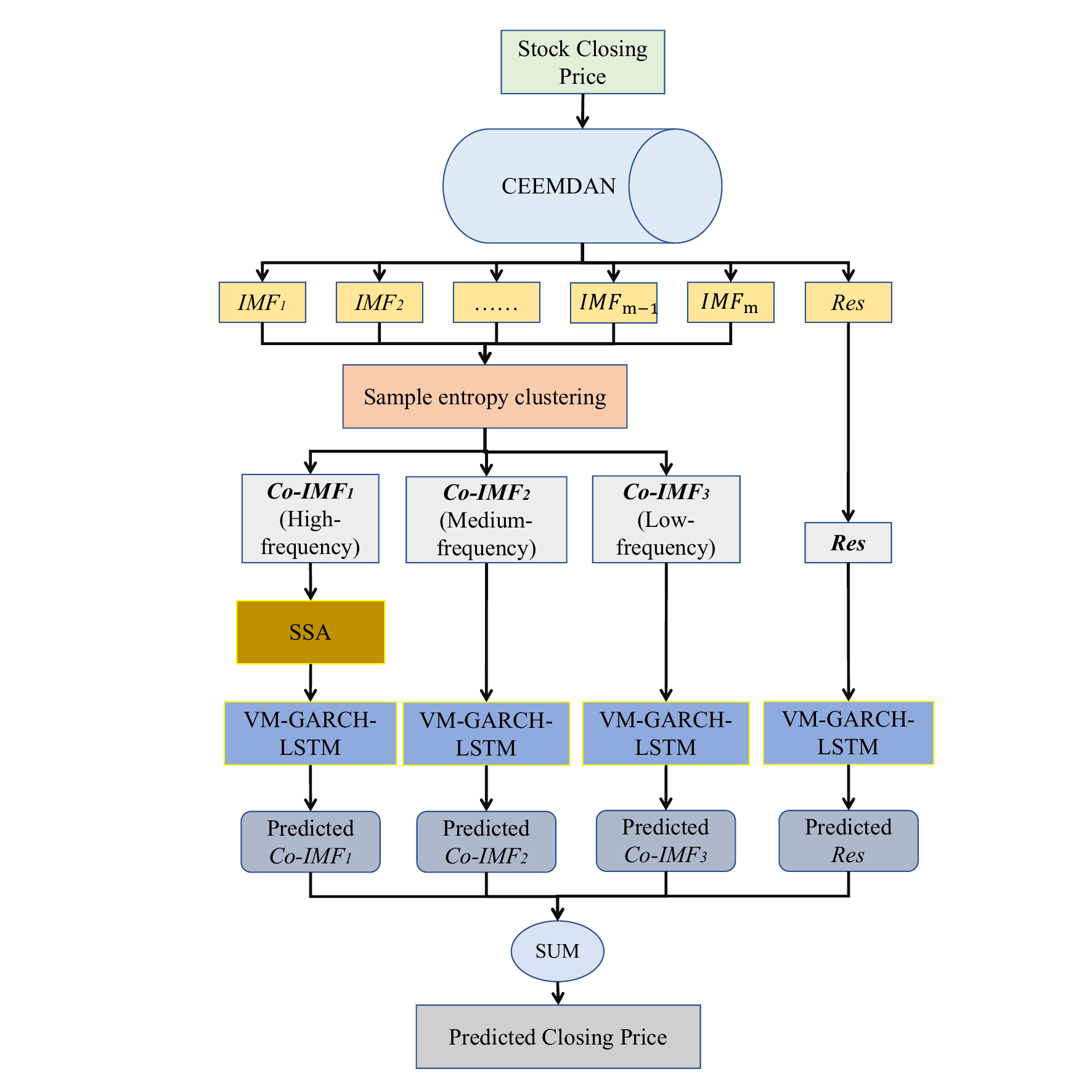}
	\caption{Flowchart of the proposed CEEMDAN-SE-SSA-VMGL model.}
	\label{FIG:CSVMGL}
\end{figure}

This section describes the complete methodology of the proposed CEEMDAN-SE-SSA-VMGL hybrid framework, and Figure~\ref{FIG:CSVMGL} shows its overall flowchart. The framework integrates three components: data-side decomposition and clustering, signal-side singular spectrum analysis for the alleviation of mode mixing, and model-side volatility-adaptive gated prediction. The presentation proceeds in the natural order of the data flow. First, the basic principles of CEEMDAN (complete ensemble empirical mode decomposition with adaptive noise), the complexity measure of sample entropy, and the component-organization strategy of K-means clustering are introduced. The singular spectrum analysis adopted for the mode-mixing problem of high-frequency components is then described. Finally, the design of the volatility-modulated gated long short-term memory network proposed in this study is detailed, including the differentiable GARCH risk gate, the volatility-driven dynamic-scaling mechanism of the forget gate, and the implementation of the end-to-end joint optimization.

\subsection{Complete ensemble empirical mode decomposition with adaptive noise (CEEMDAN)}
Empirical mode decomposition (EMD) is prone to mode mixing, and ensemble EMD (EEMD) alleviates this problem at the cost of residual noise and reconstruction error. To overcome these drawbacks, \citet{Torres2011} proposed a complete ensemble EMD with adaptive noise (CEEMDAN). CEEMDAN injects adaptive noise, that is, white-noise IMF components of a specific order, at each decomposition stage for noise-assisted analysis. The approach guarantees exact reconstruction of the original signal and suppresses mode mixing and residual noise through an adaptive noise-subtraction mechanism \citep{Shen2026}. Its mathematical procedure can be summarized as follows.

Let the original time series to be decomposed be $X(t)$. First, $J$ sequences $W_j(t)\ (j=1,2,\dots,J)$ of Gaussian white noise following the standard normal distribution are generated independently, and the initial noise-strength coefficient $\lambda_0$ is introduced to construct $J$ noise-added auxiliary sequences:
\begin{equation}
	X_j(t)=X(t)+\lambda_0 W_j(t),\quad j=1,2,\dots,J
\end{equation}

Empirical mode decomposition is performed on each auxiliary sequence $X_j(t)$. Let the operator $\mathcal{E}_k[\cdot]$ denote the operation that performs EMD on the input signal and extracts its $k$-th-order intrinsic mode function. Ensemble-averaging the first-order components of the noise-added sequences yields the first intrinsic mode of the CEEMDAN decomposition:
\begin{equation}
	\mathrm{IMF}_1(t)=\frac{1}{J}\sum_{j=1}^{J}\mathcal{E}_1\big[X_j(t)\big]
\end{equation}
Removing this first-order mode from the original series gives the first-level residual:
\begin{equation}
	R_1(t)=X(t)-\mathrm{IMF}_1(t)
\end{equation}

Based on the first-level residual $R_1(t)$, the second noise-strength coefficient $\lambda_1$ is introduced. The first-order modes $\mathcal{E}_1[W_j(t)]$ obtained by performing EMD on each noise sequence are scaled by $\lambda_1$ and added to the residual, forming new auxiliary sequences from which the first-order EMD is again extracted; ensemble-averaging yields the second intrinsic mode:
\begin{equation}
	\mathrm{IMF}_2(t)=\frac{1}{J}\sum_{j=1}^{J}\mathcal{E}_1\Big[R_1(t)+\lambda_1\mathcal{E}_1\big[W_j(t)\big]\Big]
\end{equation}
The second-level residual is updated:
\begin{equation}
	R_2(t)=R_1(t)-\mathrm{IMF}_2(t)
\end{equation}

By analogy, after the first $k$ modes have been extracted, the $k$-th-level residual can be defined recursively as
\begin{equation}
	R_k(t)=R_{k-1}(t)-\mathrm{IMF}_k(t)
\end{equation}
and the $(k+1)$-th intrinsic mode is then constructed as follows:
\begin{equation}
	\mathrm{IMF}_{k+1}(t)=\frac{1}{J}\sum_{j=1}^{J}\mathcal{E}_1\Big[R_k(t)+\lambda_k\mathcal{E}_k\big[W_j(t)\big]\Big]
\end{equation}
where $\lambda_k$ is the adaptive noise-strength coefficient at the $k$-th stage, and $\mathcal{E}_k[W_j(t)]$ denotes the $k$-th-order IMF obtained by performing EMD on the $j$-th noise sequence $W_j(t)$. At every iteration, the operator $\mathcal{E}_1[\cdot]$ performs first-order extraction on the sum of the residual and the scaled noise component, whereas the noise component itself is taken as the IMF of the corresponding order $k$. In this way, the injected noise is adaptively matched to the residual at each stage.

The residual-update and mode-extraction processes are repeated until the resulting residual $R_k(t)$ satisfies the stopping criterion (e.g., the number of extrema is fewer than two, or the residual degenerates into a monotonic trend term) and no longer qualifies for intrinsic mode extraction. Suppose a total of $M$ intrinsic modes are finally extracted; the complete reconstruction of the original series is then
\begin{equation}
	X(t)=\sum_{k=1}^{M}\mathrm{IMF}_k(t)+R(t)
\end{equation}
where $R(t)$ is the final residual term, characterizing the long-term trend or the DC component of the sequence.

\subsection{Sample entropy (SE)}
\label{sec:sampen}

Sample entropy is a statistical measure proposed by \citet{Richman2000} for quantifying the complexity of time-series data. Compared with approximate entropy, sample entropy provides more consistent estimates and is less dependent on the record length, which makes it reliable for time series of moderate length. For a one-dimensional observation sequence $\{z_\ell\}_{\ell=1}^{\mathcal{N}}$, the computation proceeds as follows.

First, $\mathcal{Q}$-dimensional template vectors are constructed from the original sequence:
\begin{equation}
	\mathbf{Z}_\mathcal{Q}(i) = \bigl[z_i, z_{i+1}, \ldots, z_{i+\mathcal{Q}-1}\bigr], \quad 1 \le i \le \mathcal{N}-\mathcal{Q}
\end{equation}
where each vector represents the $\mathcal{Q}$ consecutive sampled values starting from the $i$-th observation point. The total number of template vectors is $\mathcal{N}-\mathcal{Q}$, which ensures consistency with the subsequent $(\mathcal{Q}+1)$-dimensional matching statistics.

Define the Chebyshev distance between vectors $\mathbf{Z}_\mathcal{Q}(i)$ and $\mathbf{Z}_\mathcal{Q}(j)$ as
\begin{equation}
	\rho\bigl[\mathbf{Z}_\mathcal{Q}(i), \mathbf{Z}_\mathcal{Q}(j)\bigr] = \max_{k=0,\ldots,\mathcal{Q}-1} \bigl|z_{i+k} - z_{j+k}\bigr|
\end{equation}

Given a pre-specified similarity tolerance $\vartheta$, for each $\mathbf{Z}_\mathcal{Q}(i)$ we count the number of indices $j$ (where $1 \le j \le \mathcal{N}-\mathcal{Q}$ and $j \neq i$) satisfying $\rho\bigl[\mathbf{Z}_\mathcal{Q}(i), \mathbf{Z}_\mathcal{Q}(j)\bigr] \le \vartheta$, and denote it by $\mathfrak{B}_i$. Define
\begin{equation}
	\mathfrak{b}_i^{(\mathcal{Q})}(\vartheta) = \frac{1}{\mathcal{N}-\mathcal{Q}-1}\mathfrak{B}_i, \qquad \mathfrak{B}^{(\mathcal{Q})}(\vartheta) = \frac{1}{\mathcal{N}-\mathcal{Q}}\sum_{i=1}^{\mathcal{N}-\mathcal{Q}} \mathfrak{b}_i^{(\mathcal{Q})}(\vartheta)
\end{equation}

Similarly, construct $(\mathcal{Q}+1)$-dimensional vectors $\mathbf{Z}_{\mathcal{Q}+1}(i)$ with the index range $1 \le i \le \mathcal{N}-\mathcal{Q}$:
\begin{equation}
	\mathbf{Z}_{\mathcal{Q}+1}(i) = \bigl[z_i, z_{i+1}, \ldots, z_{i+\mathcal{Q}}\bigr]
\end{equation}
For each $\mathbf{Z}_{\mathcal{Q}+1}(i)$, count the number of indices $j$ (where $1 \le j \le \mathcal{N}-\mathcal{Q}$ and $j \neq i$) satisfying the distance condition, and denote it by $\mathfrak{A}_i$. Define
\begin{equation}
	\mathfrak{a}_i^{(\mathcal{Q})}(\vartheta) = \frac{1}{\mathcal{N}-\mathcal{Q}-1}\mathfrak{A}_i, \qquad \mathfrak{A}^{(\mathcal{Q})}(\vartheta) = \frac{1}{\mathcal{N}-\mathcal{Q}}\sum_{i=1}^{\mathcal{N}-\mathcal{Q}} \mathfrak{a}_i^{(\mathcal{Q})}(\vartheta)
\end{equation}

Here, $\mathfrak{B}^{(\mathcal{Q})}(\vartheta)$ represents the conditional probability that two subsequences match $\mathcal{Q}$ points within the tolerance $\vartheta$, whereas $\mathfrak{A}^{(\mathcal{Q})}(\vartheta)$ represents the probability of matching $\mathcal{Q}+1$ points. The sample entropy is defined as
\begin{equation}
	\mathrm{SE}(\mathcal{Q}, \vartheta) = -\ln\left[\frac{\mathfrak{A}^{(\mathcal{Q})}(\vartheta)}{\mathfrak{B}^{(\mathcal{Q})}(\vartheta)}\right]
\end{equation}

A smaller sample-entropy value indicates higher self-similarity and lower complexity of the sequence, whereas a larger value indicates stronger irregularity and more pronounced randomness. In the proposed framework, sample entropy serves as the complexity measure for subsequence classification: subsequences with similar entropy values are grouped into the same reconstruction cluster, which facilitates the extraction of multi-scale features from the raw data.

\subsection{K-means clustering}
\label{sec:kmeans}
K-means is a centroid-based partitioning algorithm that iteratively minimizes the within-cluster sum of squared errors to divide $\mathcal{N}$ unlabeled observations into $\mathcal{K}$ disjoint clusters \citep{McQueen1967}:

\begin{equation}
\mathrm{SSE} = \sum_{k=1}^{\mathcal{K}} \sum_{\mathbf{x}_\ell \in \mathcal{C}_k} \bigl\|\mathbf{x}_\ell - \boldsymbol{\mu}_k\bigr\|_2^2 = \sum_{k=1}^{\mathcal{K}} \sum_{\mathbf{x}_\ell \in \mathcal{C}_k} \sum_{d=1}^{\mathcal{D}} \bigl(x_{\ell,d} - \mu_{k,d}\bigr)^2
\end{equation}

where $\mathcal{N}$ and $\mathcal{K}$ denote the sample size and the number of clusters, respectively; $\mathbf{x}_\ell \in \mathbb{R}^{\mathcal{D}}$ is the $\ell$-th observation, $\mathcal{C}_k$ is the $k$-th cluster with centroid $\boldsymbol{\mu}_k = \frac{1}{|\mathcal{C}_k|}\sum_{\mathbf{x}_\ell \in \mathcal{C}_k} \mathbf{x}_\ell$, and $\|\cdot\|_2$ denotes the Euclidean norm.

\subsection{Singular Spectrum Analysis (SSA)}
\label{sec:ssa}

Singular spectrum analysis (SSA) extracts the components of a nonlinear time series by decomposing and reconstructing its trajectory matrix, thereby separating the signal from the noise of the original series \citep{Broomhead1986}; beyond this decomposition, it also reveals the important patterns and trends hidden in the data \citep{Emmanuel2022}. Its computational procedure comprises four consecutive stages: embedding, spectral decomposition, grouping or clustering, and diagonal averaging.

Given an observed sequence $\{s_\tau\}_{\tau=1}^{\mathcal{T}}$, first choose an embedding dimension $\mathcal{W}$ satisfying $2 \le \mathcal{W} \le \mathcal{T}/2$, and define $\mathcal{V} = \mathcal{T} - \mathcal{W} + 1$. The trajectory matrix $\mathbf{S} \in \mathbb{R}^{\mathcal{W} \times \mathcal{V}}$ is constructed by delay embedding:

\begin{equation}
	\mathbf{S} =
	\begin{pmatrix}
		s_1 & s_2 & \cdots & s_{\mathcal{V}} \\
		s_2 & s_3 & \cdots & s_{\mathcal{V}+1} \\
		\vdots & \vdots & \ddots & \vdots \\
		s_{\mathcal{W}} & s_{\mathcal{W}+1} & \cdots & s_{\mathcal{T}}
	\end{pmatrix}
\end{equation}

This matrix with Hankel structure converts temporal analysis into matrix analysis, each column representing a delay-shifted copy of the original sequence.

The trajectory matrix $\mathbf{S}$ is then subjected to singular value decomposition (SVD):

\begin{equation}
	\mathbf{S} = \mathbf{\Upsilon} \mathbf{\Omega} \mathbf{\Gamma}^{\mathrm{T}}
\end{equation}

where $\mathbf{\Upsilon} = [\boldsymbol{\upsilon}_1, \boldsymbol{\upsilon}_2, \ldots, \boldsymbol{\upsilon}_{\mathcal{W}}] \in \mathbb{R}^{\mathcal{W} \times \mathcal{W}}$ and $\mathbf{\Gamma} = [\boldsymbol{\gamma}_1, \boldsymbol{\gamma}_2, \ldots, \boldsymbol{\gamma}_{\mathcal{V}}] \in \mathbb{R}^{\mathcal{V} \times \mathcal{V}}$ are orthogonal matrices, and $\mathbf{\Omega} \in \mathbb{R}^{\mathcal{W} \times \mathcal{V}}$ is a rectangular diagonal matrix whose non-negative diagonal elements $\omega_1 \ge \omega_2 \ge \cdots \ge \omega_{\mathcal{W}} \ge 0$ are called the singular values. Equivalently, eigendecomposition of the covariance matrix $\mathbf{C} = \mathbf{S}\mathbf{S}^{\mathrm{T}}$ yields eigenvalues $\chi_m = \omega_m^2$ and the corresponding eigenvectors $\boldsymbol{\upsilon}_m$; the trajectory matrix can be expanded as

\begin{equation}
	\mathbf{S} = \sum_{m=1}^{\mathcal{W}} \omega_m \boldsymbol{\upsilon}_m \boldsymbol{\gamma}_m^{\mathrm{T}} = \sum_{m=1}^{\mathcal{W}} \sqrt{\chi_m} \, \boldsymbol{\upsilon}_m \boldsymbol{\gamma}_m^{\mathrm{T}}
\end{equation}

After spectral decomposition, the complete index set $\{1, 2, \ldots, \mathcal{W}\}$ is partitioned into $\mathcal{R}$ pairwise disjoint subsets $\mathcal{J}_1, \mathcal{J}_2, \ldots, \mathcal{J}_{\mathcal{R}}$ satisfying $\bigcup_{r=1}^{\mathcal{R}} \mathcal{J}_r = \{1, 2, \ldots, \mathcal{W}\}$ and $\mathcal{J}_r \cap \mathcal{J}_{r'} = \emptyset$ for any $r \neq r'$. Let $\mathcal{J} = \{j_1, j_2, \ldots, j_p\}$ be any such subset; the corresponding grouped matrix is defined as

\begin{equation}
	\mathbf{S}_{\mathcal{J}} = \sum_{m \in \mathcal{J}} \omega_m \boldsymbol{\upsilon}_m \boldsymbol{\gamma}_m^{\mathrm{T}}
\end{equation}

Thus the trajectory matrix is decomposed exactly as the sum of all grouped matrices, $\mathbf{S} = \sum_{r=1}^{\mathcal{R}} \mathbf{S}_{\mathcal{J}_r}$. The grouping strategy, usually based on the elbow of the singular-value spectrum or on the eigenvalue contribution rate, assigns the trend, periodic, and noise components to different subsets.

Finally, each grouped matrix $\mathbf{S}_{\mathcal{J}_r}$ is converted back to a one-dimensional sequence of length $\mathcal{T}$ by diagonal averaging (Hankelization). Let $\mathbf{M}$ be any $\mathcal{W} \times \mathcal{V}$ grouped matrix with elements $m_{ij}$. Define $\mathcal{W}^* = \min(\mathcal{W}, \mathcal{V})$ and $\mathcal{V}^* = \max(\mathcal{W}, \mathcal{V})$, and construct an auxiliary matrix $\mathbf{M}^*$:

\begin{equation}
	m^*_{ij} =
	\begin{cases}
		m_{ij}, & \mathcal{W} \le \mathcal{V} \\[4pt]
		m_{ji}, & \mathcal{W} > \mathcal{V}
	\end{cases}
\end{equation}

Applying diagonal averaging to $\mathbf{M}^*$ yields the sequence $\{d_1, d_2, \ldots, d_{\mathcal{T}}\}$:

\begin{equation}
	d_k =
	\begin{cases}
		\dfrac{1}{k} \displaystyle\sum_{i=1}^{k} m^*_{i, k-i+1}, & 1 \le k < \mathcal{W}^* \\[14pt]
		\dfrac{1}{\mathcal{W}^*} \displaystyle\sum_{i=1}^{\mathcal{W}^*} m^*_{i, k-i+1}, & \mathcal{W}^* \le k \le \mathcal{V}^* \\[14pt]
		\dfrac{1}{\mathcal{T}-k+1} \displaystyle\sum_{i=k-\mathcal{V}^*+1}^{\mathcal{W}^*} m^*_{i, k-i+1}, & \mathcal{V}^* < k \le \mathcal{T}
	\end{cases}
\end{equation}

After these four stages, the original series is decomposed into $\mathcal{R}$ subsequences at different time scales, which separates the trend, periodic, and noise components of the non-stationary financial time series.

\subsection{Long Short-Term Memory (LSTM)}
The long short-term memory (LSTM) network is an extension of the recurrent neural network (RNN). Its gating mechanism resolves, to a certain extent, the vanishing- or exploding-gradient problems that traditional RNNs face on long sequences \citep{Hochreiter1997}, which makes it well suited to sequence data with long-range dependencies. Figure~\ref{FIG:LSTM} shows the structure of an LSTM unit, and the output of an LSTM unit can be formulated as follows:
\begin{figure}
  \centering
   \includegraphics[width=0.65\textwidth]{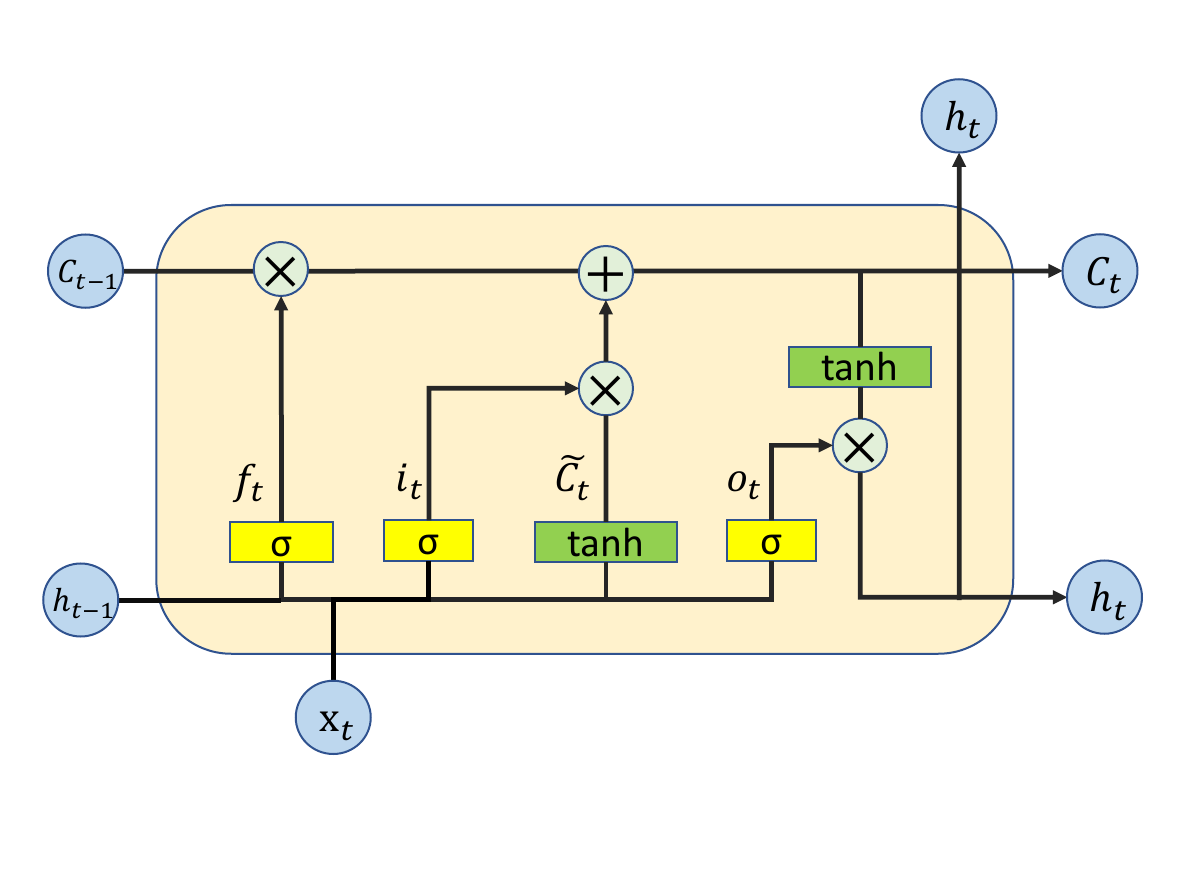}
    \caption{Schematic diagram of LSTM structure.}
    \label{FIG:LSTM}
\end{figure}
\begin{align}
	C_t &= f_t \odot C_{t-1} + i_t \odot \tilde{C}_t \\
	f_t &= \sigma(W_f \cdot [h_{t-1}, x_t] + b_f) \label{eq:4} \\
	i_t &= \sigma(W_i \cdot [h_{t-1}, x_t] + b_i) \\
	o_t &= \sigma(W_o \cdot [h_{t-1}, x_t] + b_o)\\
	\tilde{C}_t& = \tanh(W_c \cdot [h_{t-1}, x_t] + b_c)\\
	h_t &= o_t \odot \tanh({C}_t)
\end{align}
where $\odot$ denotes element-wise multiplication and $[h_{t-1}, x_t]$ denotes vector concatenation. The forget gate $f_t$ controls the retention ratio of the historical state $C_{t-1}$, the input gate $i_t$ regulates the injection of the new candidate state $\tilde{C}_t$, and the output gate $o_t$ determines the output of the current hidden state $h_t$. $W$ and $b$ denote the weight matrices and biases, respectively, and $\sigma(\cdot)$ denotes the sigmoid activation function. All cell states maintain the same dimension to ensure computational compatibility.

\subsection{Volatility-Modulated GARCH-LSTM (VMGL)}
\label{sec:vmgl}
Although the gating mechanism of the standard LSTM alleviates the gradient problem on long sequences, the network processes all input information without differentiation: noise of different magnitudes in a financial series is memorized equally, which leads to unstable prediction in high-volatility periods. To address this limitation, a volatility-modulated GARCH-LSTM (VMGL) is proposed. VMGL embeds a differentiable GARCH layer into the LSTM cell as a unified risk-gate module, so that the volatility estimate and the prediction network are optimized jointly in an end-to-end manner.
\begin{figure}
  \centering
   \includegraphics[width=0.65\textwidth]{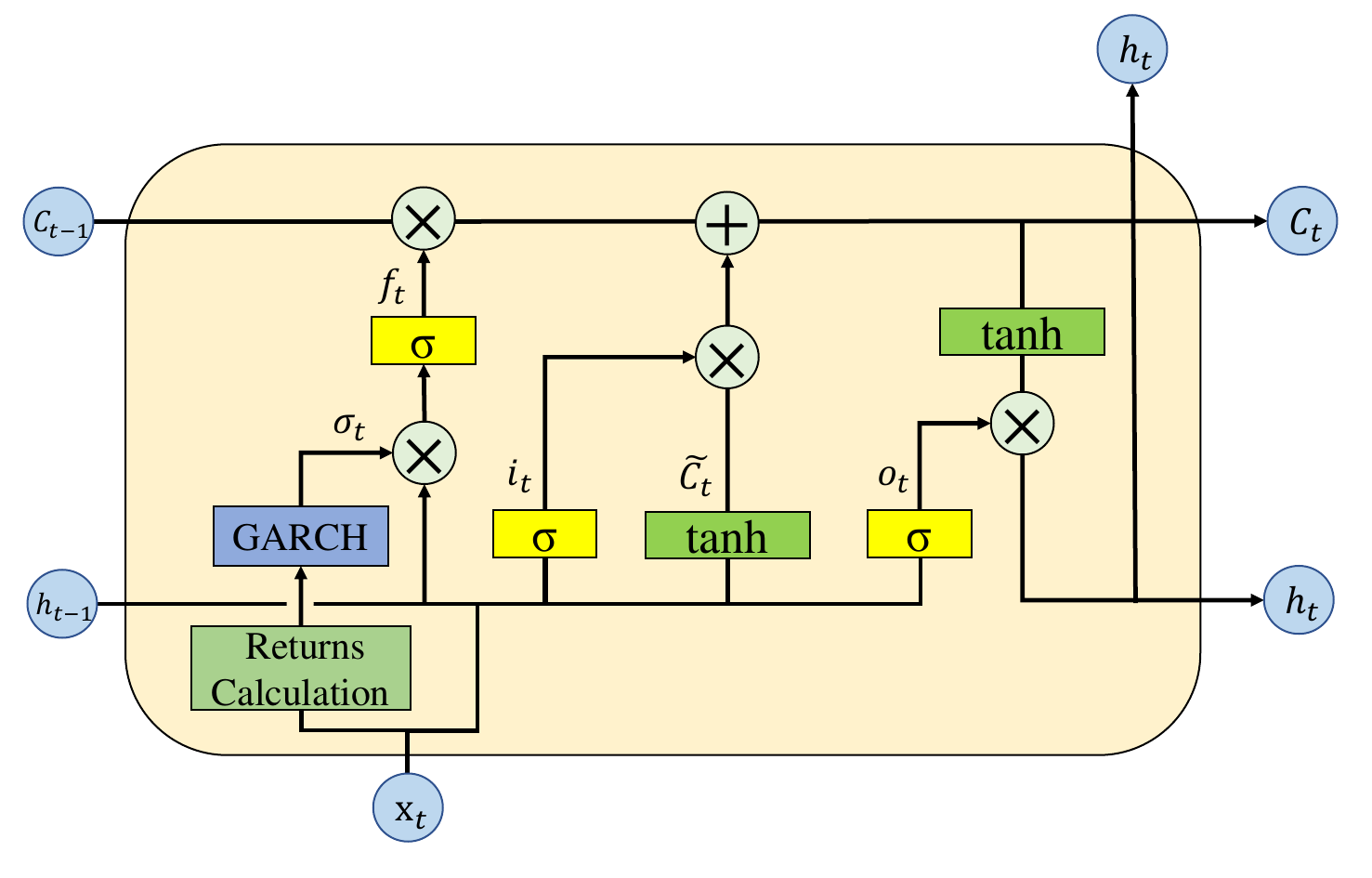}
    \caption{Schematic diagram of VMGL structure.}
    \label{FIG:VMGL}
\end{figure}
The risk-gate module adopts a dual-channel input structure. The first channel receives the normalized price series, which enters the LSTM cell as the main input. The second channel receives the raw price series and feeds it to a return-computation layer, which extracts the log return $r_t = \log P_t - \log P_{t-1}$, where $P_t$ is the raw price at the current time step. This return series is then the sole input to the differentiable GARCH layer.

The differentiable GARCH layer performs the GARCH(1,1) recurrence to estimate the time-varying conditional variance. To simplify volatility modeling and focus on the conditional heteroskedasticity structure, this paper assumes a zero conditional mean for the returns, i.e., $\mu_t = 0$, a common setting in financial volatility forecasting. The recurrence is
\begin{equation}
\sigma_t^2 = \omega + \alpha \epsilon_{t-1}^2 + \beta \sigma_{t-1}^2, \quad
\epsilon_{t-1} = r_{t-1},
\end{equation}
where $\sigma_t^2$ is the conditional variance and $\epsilon_{t-1}$ is the residual. In this GARCH layer, the three coefficients $\omega$, $\alpha$, and $\beta$ are treated as learnable variables rather than being fixed in advance or fitted offline by traditional maximum-likelihood estimation. They are updated end-to-end through backpropagation together with the prediction loss of the whole model.

To ensure the positivity and stationarity of the GARCH coefficients, VMGL imposes two constraints on the learnable parameters. First, the Softplus function maps the unconstrained raw parameters to positive values:
\begin{equation}
\omega = \text{Softplus}(\omega_{\text{raw}}), \quad
\alpha = \text{Softplus}(\alpha_{\text{raw}}), \quad
\beta = \text{Softplus}(\beta_{\text{raw}}),
\end{equation}
where $\text{Softplus}(x)=\log(1+e^x)$ is always positive. Second, a dynamic scaling is introduced to satisfy the stationarity constraint $\alpha+\beta<1$:
\begin{equation}
\text{scale} = \min\left(1,\; \frac{0.99}{\alpha+\beta+\epsilon}\right), \quad
\alpha_{\text{con}} = \alpha \cdot \text{scale}, \quad
\beta_{\text{con}} = \beta \cdot \text{scale}.
\end{equation}
These constraints keep the recurrence numerically stable while allowing unconstrained gradient-descent optimization. Because the learnable GARCH parameters are updated entirely by the prediction error, the mismatch between the volatility estimate and the prediction objective that occurs in traditional two-stage methods is avoided.

VMGL scales the standard LSTM forget gate by the conditional volatility estimated in the differentiable GARCH layer:
\begin{equation}
f_t = \frac{1}{1 + \sigma_t + \epsilon} \cdot \sigma\big(\alpha_f \cdot (W_f x_t + U_f h_{t-1} + b_f) + b_{\text{bias}}\big),
\end{equation}
where $\alpha_f$ and $b_{\text{bias}}$ are learnable parameters and $\epsilon$ is a constant that prevents division by zero. The modulation factor $\frac{1}{1+\sigma_t}$ decays as the estimated volatility rises, which accelerates the forgetting of outdated patterns; it approaches $1$ as volatility falls, which preserves long-term dependence.

The loss function of VMGL is simply the mean squared prediction error
\begin{equation}
\mathcal{L} = \frac{1}{N}\sum_{t=1}^{N} (y_t - \hat{y}_t)^2.
\end{equation}
Letting $\delta_t = \partial \mathcal{L} / \partial \sigma_t^2$, backpropagation through the GARCH recurrence yields
\begin{equation}
\frac{\partial \mathcal{L}}{\partial \sigma_{t-1}^2} = \delta_t \cdot \beta_{\text{con}}. 
\end{equation}
Because $\beta_{\text{con}}<0.99$, the gradient neither explodes nor vanishes, effectively avoiding gradient anomalies on long sequences. At the same time, the gradients of the learnable GARCH parameters
\begin{align}
\frac{\partial \mathcal{L}}{\partial \omega} &= \sum_{t} \delta_t,  \\
\frac{\partial \mathcal{L}}{\partial \alpha} &= \sum_{t} \delta_t \epsilon_{t-1}^2,  \\
\frac{\partial \mathcal{L}}{\partial \beta} &= \sum_{t} \delta_t \sigma_{t-1}^2, 
\end{align}
all remain at a reasonable magnitude. This design keeps the convergence path of the model consistent across independent training runs and reduces the variation of prediction performance across different market cycles and extreme conditions. In addition, the parameter $\beta$ in the GARCH recurrence exerts a smoothing effect that suppresses high-frequency random perturbations, while the nonlinear modulation in the risk gate filters the impact of outliers on the forget gate. These designs strengthen the robustness of VMGL to the noise contained in the input data and support its stable performance on real financial data.

VMGL uses only a single-layer LSTM with 128 hidden units. Many time-series forecasting models require two or three stacked LSTM layers or stacked Transformer attention layers to reach comparable accuracy, whereas the parameter count of VMGL is only one-half to one-quarter of theirs. Because the GARCH recurrence involves only scalar operations, it adds $\mathcal{O}(1)$ computational complexity per time step and imposes negligible cost on the forward and backward passes. This lightweight design allows the model to converge faster during training and to consume fewer resources at inference.

Owing to its end-to-end structure, VMGL requires no offline re-estimation of volatility, because all computations are deterministic operations in the forward pass. Its single-layer LSTM combined with a scalar GARCH recurrence yields a simple and efficient computational graph. On commodity CPU hardware, a single forward pass is completed with very low latency, which suffices to match the sampling frequency of real-time financial data streams. Since the global min-max normalization parameters are fixed after training, the online phase requires only simple arithmetic operations. These characteristics make VMGL suitable for latency-sensitive scenarios such as high-frequency trading and tick-level risk control.

\subsection{Proposed framework}
The CEEMDAN-SE-SSA-VMGL framework proposed in this paper combines adaptive signal decomposition, sample-entropy-based clustering, singular spectrum analysis, and a risk-gated deep-learning method to handle the nonlinearity, non-stationarity, and volatility-clustering features of financial time series. Its workflow is as follows.

\textbf{Stage 1: Adaptive decomposition based on CEEMDAN.} Let the raw closing-price series be $P(t)$, which is first decomposed adaptively by the CEEMDAN algorithm. By injecting adaptive noise level by level and applying ensemble averaging, CEEMDAN decomposes $P(t)$ into $M$ intrinsic mode functions $\{\mathrm{IMF}_1(t),\mathrm{IMF}_2(t),\ldots,\mathrm{IMF}_M(t)\}$ and one residual term $R(t)$. Arranged by frequency from high to low, the IMFs capture the volatility patterns of the original series at different scales: high-frequency IMFs reflect market micro-noise and short-term oscillation, mid-frequency IMFs characterize medium-period price fluctuations, and the low-frequency IMFs together with the residual jointly depict the long-term trend and market fundamentals. The purpose of this step is to separate the original signal, whose dynamics are mixed, into components with relatively homogeneous dynamic behavior, so that the differentiated decomposition--reconstruction and modeling can be performed in the subsequent stages.

\textbf{Stage 2: Sample-entropy-based complexity quantification and K-means clustering aggregation.} For each intrinsic mode function $\mathrm{IMF}_k(t)$ generated in Stage 1, its sample-entropy value $\mathrm{SE}_k$ is computed (Section~\ref{sec:sampen}), which quantifies the irregularity and complexity of the component in a statistical sense. High-frequency components usually have large SE values because they exhibit strong randomness, whereas low-frequency components have small SE values because of their strong orderliness and self-similarity. Taking the SE values of the IMFs as the only feature, the K-means clustering algorithm (Section~\ref{sec:kmeans}) is applied with the number of clusters pre-set to 3. According to the relative magnitudes of the cluster SE values, the three clusters are then labeled as the high-frequency (HF), mid-frequency (MF), and low-frequency (LF) clusters, and the IMFs within each cluster are summed linearly to obtain the three aggregated components $P_{\mathrm{HF}}(t)$, $P_{\mathrm{MF}}(t)$, and $P_{\mathrm{LF}}(t)$. The residual term $R(t)$ is retained independently as a fourth component. This clustering--aggregation strategy avoids the computational redundancy and the overfitting risk of modeling the many IMFs one by one, and it ensures that the different frequency-domain components receive differentiated treatment in the subsequent stages.

\textbf{Stage 3: SSA targeted decomposition--reconstruction of the high-frequency aggregated component.} Because the high-frequency aggregated component $P_{\mathrm{HF}}(t)$ superimposes multiple high-frequency IMFs, its noise content is the densest, and modeling it directly would disturb prediction stability. This paper therefore subjects only $P_{\mathrm{HF}}(t)$ to the SSA decomposition--reconstruction described in Section~\ref{sec:ssa} and leaves the mid-frequency, low-frequency, and residual components untouched. In the SSA processing, a trajectory matrix is constructed for $P_{\mathrm{HF}}(t)$ with an appropriate embedding dimension $\mathcal{W}$ and subjected to singular value decomposition; the principal and noise components are separated according to the singular-value contribution rate, and the high-frequency component $\tilde{P}_{\mathrm{HF}}(t)$ is reconstructed by diagonal averaging. The rationale is that the high-frequency component suffers the most severe mode mixing and pseudo-oscillation and therefore most needs signal refinement, whereas the mid- and low-frequency components already possess high structural order, so that additional processing could only remove information useful for prediction. This targeted refinement ensures the data quality of each component before it enters the predictor.

\textbf{Stage 4: Independent VMGL prediction of each component.} The SSA-processed high-frequency component $\tilde{P}_{\mathrm{HF}}(t)$, the mid-frequency component $P_{\mathrm{MF}}(t)$, the low-frequency component $P_{\mathrm{LF}}(t)$, and the residual term $R(t)$ are fed respectively into four independent VMGL networks (Section~\ref{sec:vmgl}) for prediction. Each VMGL network independently performs the following operations. First, the log return is extracted from its input sequence and the time-varying conditional-volatility series is estimated recursively by the differentiable GARCH(1,1) layer. The raw series then enters the LSTM cell as the main input, while the estimated volatility is injected into the forget gate as a modulation signal. At high volatility, forgetting is accelerated to adapt quickly to market changes, whereas at low volatility, long-term dependence is preserved to maintain stable memory. The four VMGL networks share the same network structure (single-layer LSTM with 128 hidden units), but the GARCH learnable parameters ($\omega,\alpha,\beta$) and the LSTM weight matrices and biases of each network are optimized independently to match the unique volatility dynamics and temporal patterns of each component. Because the memory of each VMGL network is modulated by the conditional volatility of its own component, the same architecture adapts to components with markedly heterogeneous volatility, which avoids the competing-objectives dilemma that arises when a single non-adaptive model is trained on mixed signals.

\textbf{Stage 5: Linear superposition ensemble output.} The prediction results of the four VMGL networks are combined by simple linear superposition to obtain the final prediction of the raw closing-price series:
\begin{equation}
	\hat{P}(t+1)=\hat{P}_{\mathrm{HF}}(t+1)+\hat{P}_{\mathrm{MF}}(t+1)+\hat{P}_{\mathrm{LF}}(t+1)+\hat{R}(t+1)
\end{equation}
where $\hat{P}_{\mathrm{HF}}$, $\hat{P}_{\mathrm{MF}}$, $\hat{P}_{\mathrm{LF}}$, and $\hat{R}$ are the VMGL one-step prediction outputs of the high-, mid-, and low-frequency aggregated components and the residual term, respectively.

\begin{algorithm}[htbp]
	\SetAlgoLined
	\SetAlFnt{\small}
	\DontPrintSemicolon
	\label{Alg:1}
	\caption{CEEMDAN-SE-SSA-VMGL}\label{Alg:CEEMDAN-SSA-VMGL}
	\KwData{Close price series $S=\{P_t^{\mathrm{close}}\}_{t=1}^{n}$ and target $\{P_{t+1}^{\mathrm{close}}\}$; CEEMDAN trials $J$, noise coefficients $\{\varepsilon_j\}$; SampEn $(m,r)$; clusters $K$; SSA window $L$, retained singular values $r_{\mathrm{svd}}$; window size {\bf w}; split dates $D_{\mathrm{trn}},D_{\mathrm{tst}}$; VMGL: units $U$, dropout $p$, learning rate $\eta$}
	\KwResult{Trained CEEMDAN-SE-SSA-VMGL predictor $\hat{P}^{\mathrm{close}}$}

	\tcc{Stage 0: Temporal partitioning of the dataset}
	Partition $S$ into the training, validation and test segments $S_{\mathrm{trn}}$, $S_{\mathrm{val}}$, $S_{\mathrm{tst}}$ according to the split dates $D_{\mathrm{trn}},D_{\mathrm{tst}}$, where $S_{\mathrm{val}}$ is the last $10\%$ of $S_{\mathrm{trn}}$; each component derived in Stages 1--3 inherits this partition, and all trainable quantities (normalization parameters, network weights) are fitted on the training segment only\;

	\tcc{Stage 1: CEEMDAN decomposition}
	$r_0 \gets S$\;
	\For{$j \gets 1$ \KwTo $M$}{
		$\mathrm{IMF}_j \gets \frac{1}{J} \sum_{i=1}^{J} \mathcal{E}_1\big(r_{j-1} + \varepsilon_j \mathcal{E}_j(w^{(i)})\big)$, \quad $r_j \gets r_{j-1} - \mathrm{IMF}_j$\;
	}
	$R \gets r_M$, where $\mathcal{E}_k(\cdot)$ is the $k$-th IMF operator of EMD and $w^{(i)}$ are white-noise realizations\;

	\tcc{Stage 2: Sample entropy and K-means integration}
	\For{$j \gets 1$ \KwTo $M$}{
		$\mathrm{SE}_j \gets \mathrm{SampEn}(\mathrm{IMF}_j, m, r) = -\ln \dfrac{A_j^{m}(r)}{B_j^{m}(r)}$\;
	}
	$\{C_1,\cdots,C_K\} \gets \mathrm{KMeans}(\{\mathrm{SE}_j\}, K)$; \quad $c_k \gets \sum_{j\in C_k} \mathrm{IMF}_j$, sorted by mean entropy descending\;

	\tcc{Stage 3: SSA denoising on the single highest-entropy component}
	$\widehat{c}_1 \gets \mathrm{SSA}(c_1, L, r_{\mathrm{svd}})$ \tcp*[r]{trajectory embedding, SVD, $r_{\mathrm{svd}}$-component reconstruction, Hankelization}

	\tcc{Stage 4: Component-wise VMGL prediction}
	\For{$c \in \{\widehat{c}_1, c_2, \cdots, c_K, R\}$}{
		Scale $c$ to $[0,1]$ with $\min_c,\max_c$ computed from the training segment of $c$ only (no look-ahead in normalization)\;
		Build windowed samples $\{x_t, y_t\}$ with window size {\bf w}\;

		\tcp{Volatility: differentiable GARCH(1,1), $\omega,\alpha,\beta$ learnable, $\alpha+\beta\le 0.99$}
		\For{$t \gets 2$ \KwTo $n$}{
			$\epsilon_t \gets \mathrm{clip}\big(\log P_t - \log P_{t-1},\, \pm 0.1\big)$, \quad $\sigma_t^2 \gets \omega + \alpha\epsilon_{t-1}^2 + \beta\sigma_{t-1}^2$\;
		}

		\tcp{Volatility-modulated forget gate}
		$f_t \gets \dfrac{1}{1 + \sigma_t + \epsilon}\, \sigma\big(\alpha_f (W_f x_t + U_f h_{t-1} + b_f) + b_{\mathrm{bias}}\big)$\;
		$i_t \gets \sigma(W_i x_t + U_i h_{t-1} + b_i)$, \quad $\tilde{c}_t \gets \tanh(W_c x_t + U_c h_{t-1} + b_c)$, \quad $o_t \gets \sigma(W_o x_t + U_o h_{t-1} + b_o)$\;
		$c_t \gets f_t \odot c_{t-1} + i_t \odot \tilde{c}_t$, \quad $h_t \gets o_t \odot \tanh(c_t)$\;

		\tcp{Training and prediction}
		Train VMGL on $c_{\mathrm{trn}}$ (Adam, lr $\eta$, MSE) with early stopping on $c_{\mathrm{val}}$; predict $\hat{c}_{\mathrm{tst}}$ and denormalize\;
	}

	\tcc{Stage 5: Fusion and evaluation}
	$\hat{P}^{\mathrm{close}} \gets \sum_c \hat{c}_{\mathrm{tst}}$\;
	Evaluate $R^2$, RMSE, MAE, MAPE and directional accuracy of $\hat{P}^{\mathrm{close}}$ on the test segment\;

	\Return{$\hat{P}^{\mathrm{close}}$}
\end{algorithm}

Finally, based on the above mathematical foundations, the overall procedure of the proposed framework is summarized in Algorithm~\ref{Alg:1}. This two-stage decomposition-and-integration strategy isolates and models the multi-scale dynamics inherent in financial markets while adapting the memory mechanism to market volatility conditions. By applying specialized techniques to components with different characteristics, the framework aims to improve predictive accuracy and robustness under different market conditions.

\section{Experiment}
\label{sec4}
This section designs a series of experiments to assess the forecasting performance of the proposed CEEMDAN-SE-SSA-VMGL framework and the independent contribution of each of its components. First, the data characteristics, the evaluation-metric system, and the model hyperparameter configuration of the five representative US and Chinese financial series are introduced. The CEEMDAN decomposition and sample-entropy clustering results are then analyzed to verify the signal decomposition and component grouping. Next, through a comprehensive comparison with eleven benchmark architectures, the overall forecasting accuracy is evaluated, and the independent gains of the signal-purification module and the risk-gating module are quantified through spectral comparisons before and after SSA reconstruction and through matched ablation experiments across decomposition front-ends. Finally, statistical significance is verified with Diebold--Mariano tests and the model-confidence-set procedure, and the overall performance of the framework is further examined from three dimensions, namely robustness, deployment feasibility, and economic utility, through sliding-window sensitivity analysis, computational-cost evaluation, and portfolio backtesting trade simulation.

\subsection{Experiment preparation}
\subsubsection{Data preparation}
To evaluate the effectiveness of the proposed forecasting method on different stock markets, this paper selects training and test data from five representative financial series: the CSI 300, CSI 500, and S\&P 500 indices, together with Apple Inc. (AAPL) and Alphabet Inc. (GOOGL). The CSI 300 index is composed of 300 leading stocks with large capitalization and good liquidity listed on the Shanghai and Shenzhen stock exchanges and reflects the overall performance of large-cap stocks in the Chinese A-share market. The CSI 500 index consists of 500 stocks that rank among the top in total market capitalization after excluding the CSI 300 constituents and the top-300 largest-capitalization stocks, representing the overall trend of Chinese mid-cap stocks. AAPL and GOOGL are representative technology companies listed on the US NASDAQ market. The S\&P 500 index covers 500 large US listed companies across multiple industries and is one of the most representative large-cap US equity benchmark indices, possessing high liquidity and comprehensively reflecting the overall performance of the US stock market.

In this experiment, we obtain historical data for the five representative financial series through the Python \texttt{efinance} library, namely the CSI 300 index (000300.SH), the CSI 500 index (000905.SH), Apple Inc. (AAPL), Alphabet Inc. (GOOGL), and the S\&P 500. The data span from January 4, 2010 to December 29, 2023, at daily frequency. This period covers a broad range of market volatility, including the 2011 European sovereign-debt crisis, the market turmoil triggered by the COVID-19 pandemic, and the subsequent recovery. Such a span is sufficient for robust model evaluation under different economic conditions. The main feature of the dataset is the daily closing price. AAPL and GOOGL use forward-adjusted prices, so some early closing prices of AAPL take negative values. To forecast the next-day closing price, we adopt a sliding-window approach that uses the closing prices of the past 10 days as input features. The whole dataset is partitioned as follows: the data from 2010 to 2022 constitute the training period, with the first 90\% used as the training set and the last 10\% as the validation set; the 2023 data serve as the test set, ensuring no temporal overlap. Before feeding data into the models, all features are scaled to the range 0 to 1 by min-max normalization. The scaling parameters (minimum and maximum) are computed on the training set only and then applied to the validation and test sets to prevent data leakage. This normalization also maps possible negative closing prices into the specified range, preserving processing consistency.

Table~\ref{tab:Descriptive_stats} presents a set of descriptive statistics for the five stock datasets, including the sample count, the quantiles (25\%, 50\%, and 75\%), mean, minimum, maximum, variance, standard deviation, skewness, and kurtosis. On this basis, the paper further tests the heteroskedasticity of the log returns of each series to verify the existence of volatility clustering. Table~\ref{tab:Heteroscedasticity} reports the results of the ARCH LM and Ljung--Box tests. The ARCH LM statistic is significant at the 1\% significance level on all datasets ($p<1.00\times10^{-4}$), strongly rejecting the null hypothesis of no ARCH effects; the Ljung--Box statistic is likewise significant at the 1\% level, indicating significant autocorrelation in the squared residuals. These results consistently confirm significant heteroskedasticity in all five series, providing empirical justification for the subsequent use of GARCH-type models for volatility modeling.

\begin{table}[htbp]
	\centering
	\footnotesize
	\begin{threeparttable}
	\caption{Descriptive statistics of the five datasets.}
	\label{tab:Descriptive_stats}
	\begin{tabular}{lccccc}
		\toprule
		{\textbf{Metric}} & {\textbf{CSI300}} & {\textbf{CSI500}} & {\textbf{AAPL}} & {\textbf{GOOGL}} & {\textbf{S\&P500}} \\
		\midrule
		Sample Count & 3401 & 3401 & 3522 & 3521 & 3521 \\
		Mean & 3493.11 & 5403.82 & 52.59 & 52.68 & 221.78 \\
		Min & 2086.97 & 2749.13 & $-3.17$ & 9.47 & 72.26 \\
		Max & 5807.72 & 11545.89 & 195.83 & 148.41 & 465.31 \\
		25\% & 2850.30 & 4368.61 & 10.07 & 20.50 & 124.00 \\
		50\% & 3463.64 & 5549.67 & 23.66 & 39.58 & 193.95 \\
		75\% & 3990.66 & 6311.70 & 85.79 & 72.53 & 300.29 \\
		Variance & 680091.25 & 1679273.30 & 3331.86 & 1529.61 & 12796.19 \\
		Std. Dev. & 824.68 & 1295.87 & 57.72 & 39.11 & 113.12 \\
		Skewness & 0.28 & 0.38 & 1.07 & 0.91 & 0.53 \\
		Kurtosis & $-0.55$ & 0.83 & $-0.37$ & $-0.36$ & $-0.98$ \\
		\bottomrule
	\end{tabular}
	\begin{tablenotes}[flushleft]
		\footnotesize
		\item \textbf{Note:} Daily closing prices from January 4, 2010 to December 29, 2023. The training set (2010--2022) is split into training (90\%) and validation (10\%) subsets, with the validation set taken from the last 10\% of the training period. The test set covers 2023. AAPL and GOOGL are forward-adjusted.
	\end{tablenotes}
	\end{threeparttable}
\end{table}

\begin{table}[htbp]
	\centering
	\footnotesize
	\begin{threeparttable}
	\caption{Heteroscedasticity test results.}
	\label{tab:Heteroscedasticity}
	\begin{tabular}{lccccc}
		\toprule
		{\textbf{Metric}} & {\textbf{CSI300}} & {\textbf{CSI500}} & {\textbf{AAPL}} & {\textbf{GOOGL}} & {\textbf{S\&P500}} \\
		\midrule
		Sample Count & 3401 & 3401 & 3522 & 3521 & 3521 \\
		ARCH LM Statistic & 361.789 & 527.848 & 1047.014 & 95.469 & 1354.502 \\
		ARCH LM p-value & $1.25\times10^{-71}$\textsuperscript{***} & $4.92\times10^{-107}$\textsuperscript{***} & $1.39\times10^{-218}$\textsuperscript{***} & $4.38\times10^{-16}$\textsuperscript{***} & $6.59\times10^{-285}$\textsuperscript{***} \\
		Ljung--Box Statistic & 859.26 & 1451.08 & 3135.80 & 155.02 & 4571.61 \\
		Ljung--Box p-value & $3.72\times10^{-178}$\textsuperscript{***} & $9.29\times10^{-306}$\textsuperscript{***} & $<1.00\times10^{-4}$\textsuperscript{***} & $3.46\times10^{-28}$\textsuperscript{***} & $<1.00\times10^{-4}$\textsuperscript{***} \\
		Heteroscedasticity & Yes & Yes & Yes & Yes & Yes \\
		\bottomrule
	\end{tabular}
	\begin{tablenotes}[para,flushleft]
		\small
		\item \textbf{Note:} This table presents heteroscedasticity test results for the daily returns of five financial assets over the full sample period. ARCH LM denotes the Lagrange Multiplier test statistic for ARCH effects with 10 lags; LB represents the Ljung--Box test statistic applied to squared residuals for autocorrelation in volatility. Superscripts *, **, and *** indicate statistical significance at the 10\%, 5\%, and 1\% levels, respectively. The ARCH LM and Ljung--Box statistics are significant at the 1\% level for all datasets, rejecting the null hypotheses of no ARCH effects and no autocorrelation in squared residuals. Consequently, the presence of heteroscedasticity is confirmed across all series.
	\end{tablenotes}
	\end{threeparttable}
\end{table}

\subsubsection{Evaluation metrics of forecasting models}
To compare the prediction accuracy of each forecasting model with that of the alternative architectures, this paper adopts four evaluation metrics, namely the mean absolute error (MAE), the mean absolute percentage error (MAPE), the root mean square error (RMSE), and the coefficient of determination ($R^2$). They are computed as follows:
\begin{align}
MAE &= \frac{1}{n}\sum_{i=1}^{n}\left|\hat{x}_i - x_i\right|, \label{eq:mae} \\
MAPE &= \frac{1}{n}\sum_{i=1}^{n}\left|\frac{\hat{x}_i - x_i}{x_i}\right| \times 100\%, \label{eq:mape} \\
RMSE &= \sqrt{\frac{1}{n}\sum_{i=1}^{n}(\hat{x}_i - x_i)^2}, \label{eq:rmse} \\
R^2 &= 1 - \frac{\sum_{i=1}^{n}(\hat{x}_i - x_i)^2}{\sum_{i=1}^{n}(x_i - \bar{x}_i)^2}. \label{eq:r2}
\end{align}

\subsubsection{Parameter acquisition process and settings}
In this study, the decomposition and clustering modules of the proposed CEEMDAN-SE-SSA-VMGL framework adopt fixed parameter configurations, avoiding subjective manual adjustment for each dataset and ensuring cross-dataset experimental consistency and reproducibility. Specifically, the number of noise trials in CEEMDAN is fixed at 100 and the noise standard-deviation coefficient is set to 0.2. The sample-entropy computation adopts the standard configuration of embedding dimension $m=2$ and similarity tolerance $r=0.2$. The sample-entropy-based K-means clustering uniformly divides the intrinsic mode functions into three clusters, and the random seed of the clustering results is kept consistent with the global random seed. The embedding-window length of singular spectrum analysis is fixed at 15, principal-component reconstruction retains 5 leading eigenvectors, and it is applied only to the high-frequency aggregated component identified by sample entropy, so as to balance denoising strength against signal fidelity.

On the model side, a lookback window of 10 trading days is implemented to establish a consistent forecasting horizon, i.e., the closing-price series of the previous 10 trading days is used to forecast the next-day closing price. The VMGL network adopts a single-layer long short-term memory architecture with 128 hidden units, extracting temporal features while keeping the parameter count low. The output layer uses a single neuron with linear activation to suit the regression prediction task. The forget-gate activation is adjusted by a learnable slope parameter $\alpha_f$ (initial value 1.0) and a learnable bias term (initialized to a zero vector), and is dynamically modulated by the volatility scaling factor estimated by GARCH. The initial parameters of the differentiable GARCH layer are set to $\omega_0 = 1 \times 10^{-6}$, $\alpha_0 = 0.1$, and $\beta_0 = 0.8$, and the variance process is guaranteed to be stable by the stationarity constraint $\alpha + \beta \leq 0.99$. GARCH recurses over the log returns of the input sequence, with a return truncation value of 0.1 to prevent outliers from distorting the variance estimate. A global dropout rate of 0.1 is applied to suppress overfitting.

Model training uses the Adam optimizer with a learning rate of $1 \times 10^{-3}$ and the mean squared error loss function. The batch size is 32, balancing computational efficiency and gradient stability. Training runs for at most 200 epochs, with early stopping triggered when the validation loss shows no improvement for 20 consecutive epochs, restoring the best weights. In addition, a learning-rate decay strategy is adopted: when the validation loss shows no improvement within 15 epochs, the learning rate is halved, with a minimum learning rate of $1 \times 10^{-6}$. All input data are scaled to the $[0, 1]$ interval by min-max normalization, with the normalization parameters computed from the training set only to prevent data leakage.

\subsection{Analysis of decomposition results}

\begin{figure}
  \centering
   \includegraphics[width=0.75\textwidth]{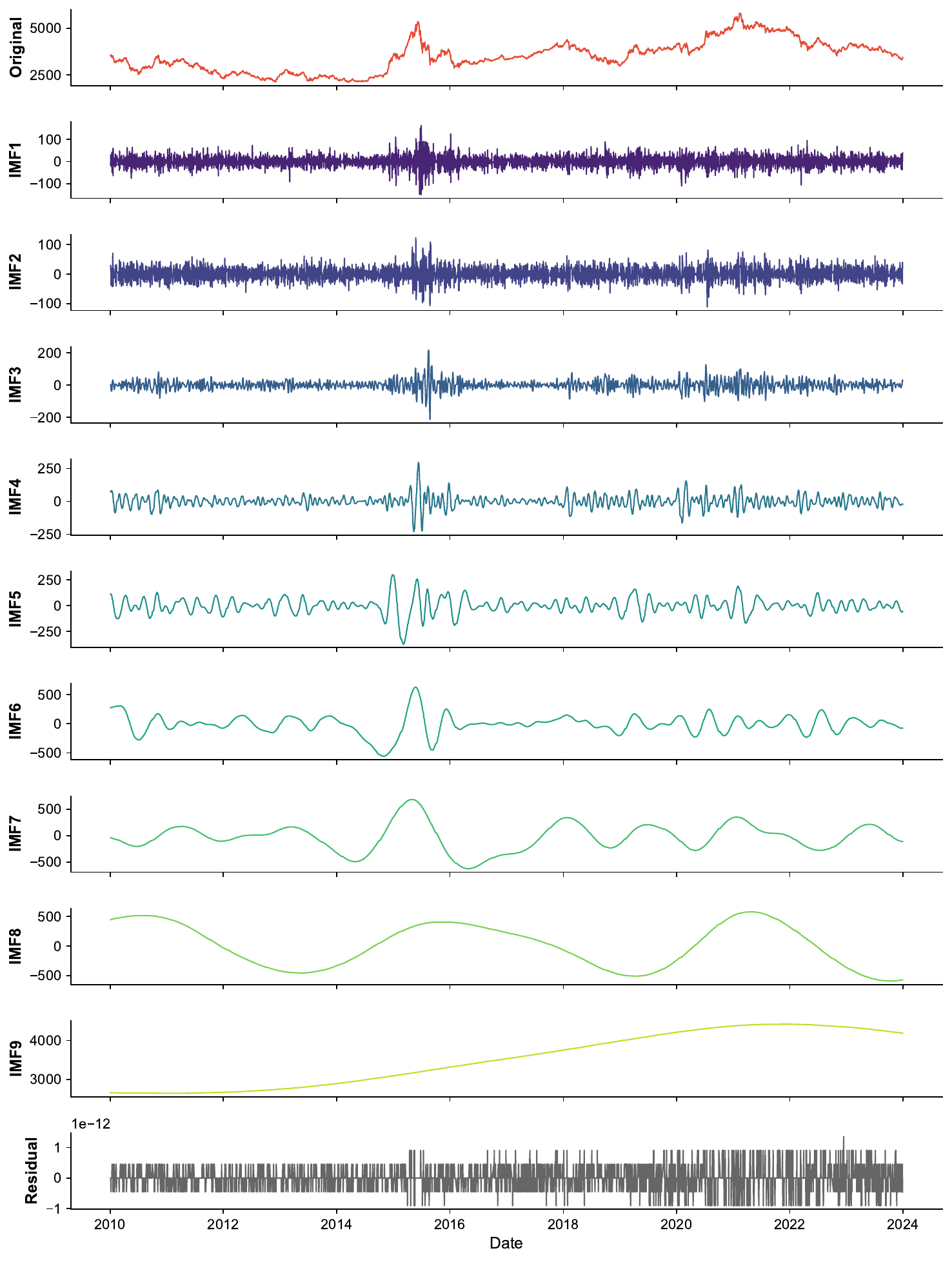}
    \caption{CEEMDAN decomposition results of the CSI 300 index.}
    \label{FIG:CEEMDAN_Decomposition}
\end{figure}

\begin{table}[htbp]
\centering
\begin{threeparttable}
\caption{Sample entropy of each sub-series from the CEEMDAN decomposition.}
\label{tab:ceemdan_imf_sampen}
\begin{tabular}{llllll}
\toprule
IMF & CSI 300 & CSI 500 & AAPL & GOOGL & S\&P 500 \\
\midrule
IMF1 & 1.846552$^{*}$ & 1.781471$^{*}$ & 2.022148$^{*}$ & 1.962428$^{*}$ & 2.019072$^{*}$ \\
IMF2 & 2.092127$^{*}$ & 2.007787$^{*}$ & 2.061346$^{*}$ & 2.039358$^{*}$ & 2.062140$^{*}$ \\
IMF3 & 0.639570$^{**}$ & 0.580142$^{**}$ & 0.587209$^{**}$ & 0.551567$^{**}$ & 0.633861$^{**}$ \\
IMF4 & 0.521673$^{**}$ & 0.303984$^{**}$ & 0.276680$^{***}$ & 0.306324$^{**}$ & 0.357532$^{**}$ \\
IMF5 & 0.293600$^{***}$ & 0.144551$^{***}$ & 0.107043$^{***}$ & 0.178634$^{***}$ & 0.158862$^{***}$ \\
IMF6 & 0.090415$^{***}$ & 0.116054$^{***}$ & 0.064036$^{***}$ & 0.127954$^{***}$ & 0.101392$^{***}$ \\
IMF7 & 0.036229$^{***}$ & 0.041741$^{***}$ & 0.019364$^{***}$ & 0.006764$^{***}$ & 0.014681$^{***}$ \\
IMF8 & 0.011891$^{***}$ & 0.012601$^{***}$ & 0.017795$^{***}$ & 0.012967$^{***}$ & 0.016869$^{***}$ \\
IMF9 & -- & 0.006432$^{***}$ & -- & -- & -- \\
Residual & 0.001972$^{+}$ & 0.001048$^{+}$ & 0.000993$^{+}$ & 0.001269$^{+}$ & 0.001341$^{+}$ \\
\bottomrule
\end{tabular}
\begin{tablenotes}[flushleft]
\small
\item \textbf{Note:} Entropy values are rounded to six decimal places. Symbols indicate K-means clustering groups: $^{*}$HF, $^{**}$MF, $^{***}$LF, $^{+}$Residual. The last IMF is always treated as the residual component and excluded from clustering. Missing entries (--) indicate indices with fewer IMFs.
\end{tablenotes}
\end{threeparttable}
\end{table}

\begin{table}[htbp]
\centering
\begin{threeparttable}
\caption{Sample entropy of the compressed HF, MF, LF, and residual (Res) components.}
\label{tab:clustered_sampen}
\begin{tabular}{lccccc}
\toprule
Component & CSI 300 & CSI 500 & AAPL & GOOGL & S\&P 500 \\
\midrule
HF & 1.765321 & 1.708424 & 1.121385 & 1.169481 & 1.441444 \\
MF & 0.606492 & 0.472504 & 0.587209 & 0.471241 & 0.559882 \\
LF & 0.054099 & 0.040794 & 0.075098 & 0.029009 & 0.046345 \\
Res & 0.001972 & 0.001048 & 0.000993 & 0.001269 & 0.001341 \\
\bottomrule
\end{tabular}
\begin{tablenotes}[flushleft]
\small
\item \textbf{Note:} Entropy values are rounded to six decimal places.
\end{tablenotes}
\end{threeparttable}
\end{table}

\begin{table}[htbp]
\centering
\begin{threeparttable}
\caption{Annualized volatility of frequency-grouped co-IMF components across five datasets.}
\label{tab:coimf_volatility}
\begin{tabular}{lccccc}
\toprule
Group & CSI 300 & CSI 500 & AAPL & GOOGL & S\&P 500 \\
\midrule
HF & 24.02 & 19.25  & 23.70 & 23.21 & 23.63 \\
MF & 15.46 & 12.84  & 16.47 & 16.35 & 13.07 \\
LF & 0.05  & 0.08   & 0.54  & 0.05  & 0.02 \\
Res & 0.00  & 0.00   & 0.00  & 0.00  & 0.00 \\
\bottomrule
\end{tabular}
\begin{tablenotes}[flushleft]
\small
\item \textbf{Note:} Annualized volatility values are expressed as percentages and rounded to two decimal places. The HF, MF, LF, and Res groups correspond to the frequency-based clustering assignments in Table~\ref{tab:ceemdan_imf_sampen}.
\end{tablenotes}
\end{threeparttable}
\end{table}

Figure~\ref{FIG:CEEMDAN_Decomposition} shows the CEEMDAN decomposition results of the CSI 300 index. The raw price series is progressively decomposed into intrinsic mode functions with decreasing frequency, and the final residual component captures the long-term trend. This hierarchical decomposition mechanism separates the volatility characteristics at different time scales, which provides the basis for the subsequent frequency-characteristic-based modeling strategy.

To quantify the complexity of each intrinsic mode function, this paper computes the sample entropy of all components in the five datasets. Table~\ref{tab:ceemdan_imf_sampen} reports the sample-entropy values of the subseries obtained from the CEEMDAN decomposition. A highly consistent pattern is observed across the five financial series. The first two intrinsic mode functions consistently display the highest sample entropy, with values between 1.78 and 2.01 on CSI 500 and reaching a maximum of 2.09 on CSI 300. This high entropy indicates that the high-frequency components contain abundant irregular fluctuations and noise; if fed directly without processing into the downstream predictor, they would readily degrade prediction stability. In contrast, the low-frequency intrinsic mode functions and the residual component have significantly lower entropy, mostly below 0.3, reflecting their smooth and highly predictable oscillatory patterns.

Based on the sample-entropy values, this paper uses the K-means clustering algorithm to divide the intrinsic mode functions into four component groups: high-frequency, mid-frequency, low-frequency, and residual. The clustering assignments, marked by superscripts in Table~\ref{tab:ceemdan_imf_sampen}, are highly consistent across the five datasets. The high-frequency group always contains the first two intrinsic mode functions, the mid-frequency group usually contains the third and fourth mode functions, and all remaining components except the residual are assigned to the low-frequency group. This robust grouping pattern confirms that sample entropy provides a reliable criterion for component classification in financial time-series analysis, generalizing across datasets.

After clustering, the intrinsic mode functions within each group are summed linearly to reconstruct compressed frequency-specific components. Table~\ref{tab:clustered_sampen} shows the sample entropy of these reconstructed components. The reconstructed components retain the original hierarchical entropy structure: the high-frequency component maintains the highest entropy, the mid-frequency component lies in the intermediate range, and the low-frequency and residual components keep the lowest entropy levels. This clear separation across the different frequency groups validates the effectiveness of the clustering strategy and indicates that the compressed components preserve the hierarchical complexity structure of the original decomposition.

To further reveal, from the volatility perspective, the heterogeneity among the different frequency groups, Table~\ref{tab:coimf_volatility} reports the annualized volatility of each grouped compressed component. Across all five datasets, the high-frequency components consistently display the highest annualized volatility, ranging from 19.25\% to 24.02\%; the mid-frequency components follow, ranging from 12.84\% to 16.47\%; and the volatility of the low-frequency components drops sharply to 0.02\%--0.54\%, differing from the high-frequency group by one to three orders of magnitude. This striking volatility disparity indicates that the market information carried by the different frequency groups is fundamentally different in statistical characteristics: the high-frequency group is dominated by short-cycle random perturbations, exhibiting strong time variation and unpredictability, whereas the low-frequency group behaves as a smooth long-term trend with virtually no instantaneous fluctuation. Yet, in existing decomposition--integration frameworks, the same predictor architecture is applied indiscriminately to all components, completely ignoring this deep difference in volatility characteristics. This observation directly motivates the need for volatility-adaptive memory-modulation mechanisms for the different components, providing a solid empirical basis for the design of the VMGL module.

\subsection{Performance comparison}
\begin{table*}[htbp]
	\centering
	\begin{threeparttable}
		\caption{Performance comparison of different models across five datasets.}
		\label{tab:model_comparison}
		\small
		\setlength{\tabcolsep}{3pt}
		\begin{tabular}{llccccc}
			\toprule
			\textbf{Metric} & \textbf{Model} & \textbf{CSI300} & \textbf{CSI500} & \textbf{AAPL} & \textbf{GOOGL} & \textbf{S\&P500} \\
			\midrule
			\multirow{12}{*}{RMSE}
			& LSTM & 36.35 & 50.99 & 2.8464 & 2.6203 & 4.2406 \\
			& VMGL & 33.69 & 49.88 & 2.3481 & 2.3308 & 3.5458 \\
			& EMD-SE-LSTM & 55.94 & 32.29 & 1.9680 & 2.3258 & 2.0771 \\
			& EMD-SE-VMGL & 25.98 & 27.13 & 1.9512 & 2.1187 & 1.9009 \\
			& EEMD-SE-LSTM & 60.58 & 127.11 & 3.3011 & 2.6965 & 6.9613 \\
			& EEMD-SE-VMGL & 23.87 & 40.94 & 1.8358 & 1.5530 & 2.7535 \\
			& VMD-SE-LSTM & 32.96 & 37.99 & 1.4931 & 1.4467 & 2.9601 \\
			& VMD-SE-VMGL & 17.72 & 34.19 & 1.0603 & 1.2681 & 2.2456 \\
			& CEEMDAN-SE-LSTM & 18.33 & 27.41 & 1.3869 & 1.3591 & 2.3543 \\
			& CEEMDAN-SE-VMGL & 17.96 & 27.04 & 1.3226 & 1.2675 & 2.0363 \\
			& CEEMDAN-SE-SSA-LSTM & 13.06 & 22.81 & 1.0749 & 1.0346 & 1.7502 \\
			& \textbf{Proposed model} & \textbf{11.91} & \textbf{19.98} & \textbf{0.8908} & \textbf{0.8472} & \textbf{1.3356} \\
			\midrule
			\multirow{12}{*}{MAE}
			& LSTM & 29.36 & 40.17 & 2.2313 & 1.9673 & 3.4516 \\
			& VMGL & 27.02 & 39.32 & 1.8420 & 1.7481 & 2.8617 \\
			& EMD-SE-LSTM & 50.58 & 26.05 & 1.6129 & 1.8004 & 1.6865 \\
			& EMD-SE-VMGL & 21.16 & 22.62 & 1.5927 & 1.6057 & 1.4948 \\
			& EEMD-SE-LSTM & 49.68 & 101.65 & 2.7244 & 2.1288 & 5.4727 \\
			& EEMD-SE-VMGL & 19.53 & 32.28 & 1.4716 & 1.2046 & 2.2425 \\
			& VMD-SE-LSTM & 27.76 & 30.10 & 1.2280 & 1.1241 & 2.5159 \\
			& VMD-SE-VMGL & 13.86 & 26.79 & 0.8729 & 0.9773 & 1.8398 \\
			& CEEMDAN-SE-LSTM & 14.15 & 22.26 & 1.0668 & 1.0489 & 1.9234 \\
			& CEEMDAN-SE-VMGL & 14.22 & 21.44 & 1.0386 & 0.9877 & 1.6607 \\
			& CEEMDAN-SE-SSA-LSTM & 10.64 & 17.95 & 0.8566 & 0.8020 & 1.3930 \\
			& \textbf{Proposed model} & \textbf{9.69} & \textbf{16.19} & \textbf{0.7180} & \textbf{0.6415} & \textbf{1.0528} \\
			\midrule
			\multirow{12}{*}{MAPE}
			& LSTM & 0.7675 & 0.6807 & 1.3081 & 1.7038 & 0.8385 \\
			& VMGL & 0.7050 & 0.6663 & 1.0807 & 1.5125 & 0.6959 \\
			& EMD-SE-LSTM & 1.3409 & 0.4420 & 0.9374 & 1.5240 & 0.4066 \\
			& EMD-SE-VMGL & 0.5557 & 0.3820 & 0.9309 & 1.3627 & 0.3620 \\
			& EEMD-SE-LSTM & 1.3011 & 1.7068 & 1.6019 & 1.8227 & 1.3176 \\
			& EEMD-SE-VMGL & 0.5100 & 0.5463 & 0.8621 & 1.0266 & 0.5404 \\
			& VMD-SE-LSTM & 0.7362 & 0.5094 & 0.7150 & 0.9499 & 0.6075 \\
			& VMD-SE-VMGL & 0.3611 & 0.4521 & 0.5044 & 0.8314 & 0.4486 \\
			& CEEMDAN-SE-LSTM & 0.3705 & 0.3777 & 0.6303 & 0.9060 & 0.4665 \\
			& CEEMDAN-SE-VMGL & 0.3724 & 0.3615 & 0.6129 & 0.8492 & 0.4041 \\
			& CEEMDAN-SE-SSA-LSTM & 0.2780 & 0.3039 & 0.5007 & 0.6896 & 0.3374 \\
			& \textbf{Proposed model} & \textbf{0.2534} & \textbf{0.2732} & \textbf{0.4233} & \textbf{0.5453} & \textbf{0.2556} \\
			\midrule
			\multirow{12}{*}{$R^2$}
			& LSTM & 0.9752 & 0.9755 & 0.9664 & 0.9706 & 0.9656 \\
			& VMGL & 0.9787 & 0.9765 & 0.9771 & 0.9767 & 0.9760 \\
			& EMD-SE-LSTM & 0.9413 & 0.9902 & 0.9821 & 0.9760 & 0.9915 \\
			& EMD-SE-VMGL & 0.9873 & 0.9931 & 0.9842 & 0.9808 & 0.9931 \\
			& EEMD-SE-LSTM & 0.9311 & 0.8477 & 0.9547 & 0.9688 & 0.9074 \\
			& EEMD-SE-VMGL & 0.9893 & 0.9842 & 0.9860 & 0.9897 & 0.9855 \\
			& VMD-SE-LSTM & 0.9796 & 0.9864 & 0.9907 & 0.9910 & 0.9833 \\
			& VMD-SE-VMGL & 0.9941 & 0.9890 & 0.9953 & 0.9931 & 0.9904 \\
			& CEEMDAN-SE-LSTM & 0.9937 & 0.9929 & 0.9920 & 0.9921 & 0.9894 \\
			& CEEMDAN-SE-VMGL & 0.9939 & 0.9931 & 0.9927 & 0.9931 & 0.9921 \\
			& CEEMDAN-SE-SSA-LSTM & 0.9968 & 0.9951 & 0.9952 & 0.9954 & 0.9941 \\
			& \textbf{Proposed model} & \textbf{0.9973} & \textbf{0.9962} & \textbf{0.9967} & \textbf{0.9969} & \textbf{0.9966} \\
			\bottomrule
		\end{tabular}
		\begin{tablenotes}[flushleft]
			\item \scriptsize \textbf{Note:} Bold values indicate the proposed model, which achieves superior performance across all metrics and datasets.
		\end{tablenotes}
	\end{threeparttable}
\end{table*}

\begin{figure}
  \centering
   \includegraphics[width=0.95\textwidth]{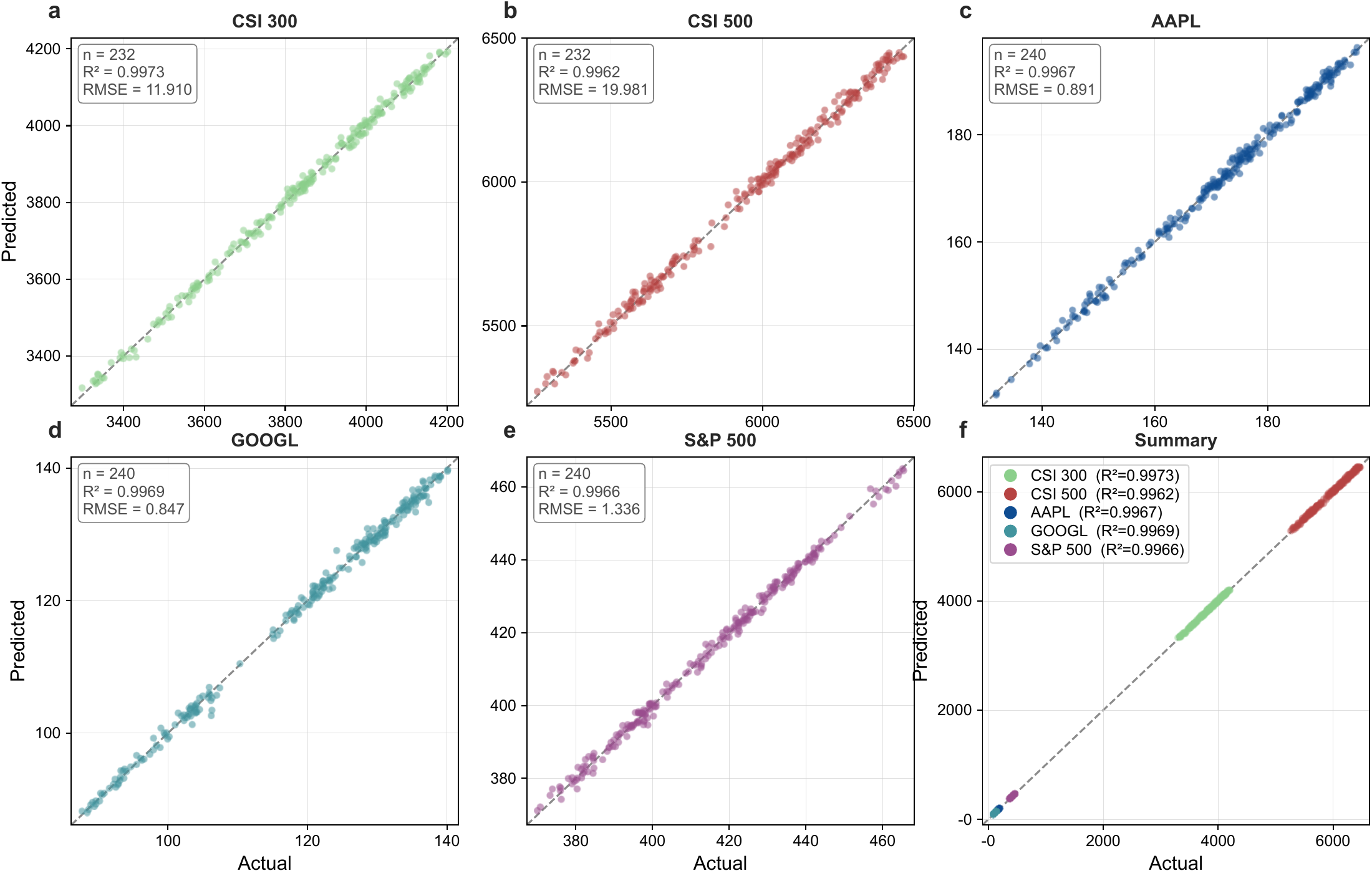}
    \caption{Scatter plots of actual versus predicted closing prices for the proposed CEEMDAN-SE-SSA-VMGL model across five datasets.}
    \label{FIG:Scatter}
\end{figure}

\begin{figure}
  \centering
   \includegraphics[width=0.95\textwidth]{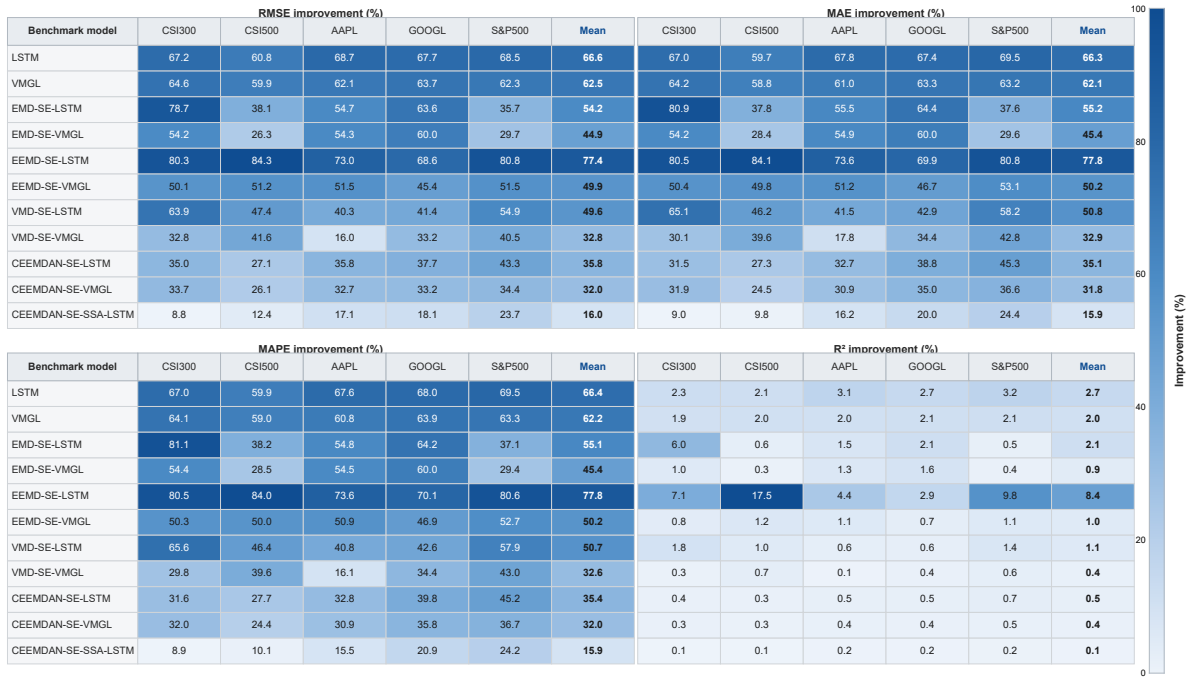}
    \caption{Percentage improvement of the proposed CEEMDAN-SE-SSA-VMGL model over eleven benchmark architectures.}
    \label{FIG:Improvements}
\end{figure}

To comprehensively validate the effectiveness and superiority of the proposed model in financial time-series forecasting, we performed an extensive comparison of the CEEMDAN-SE-SSA-VMGL model against eleven benchmark configurations. These benchmark configurations cover five decomposition strategies (none, EMD, EEMD, VMD, and CEEMDAN) combined with the LSTM and VMGL predictors, tested on five financial time series: CSI~300, CSI~500, AAPL, GOOGL, and S\&P~500. For each model--market combination, we recorded four error metrics, $R^2$, RMSE, MAE, and MAPE, and the detailed results are reported in Table~\ref{tab:model_comparison}.

CEEMDAN-SE-SSA-VMGL attains the highest coefficient of determination $R^2$ on every market, ranging from 0.9962 on CSI~500 to 0.9973 on CSI~300. Its RMSE ranges from 0.85 on GOOGL to 19.98 on CSI~500, and its MAPE stays below 0.56\% on all datasets. Among the benchmark models, CEEMDAN-SE-SSA-LSTM, which shares the same CEEMDAN-SE-SSA front-end processing module as the proposed model, is the closest-performing comparison model within the CEEMDAN family, with $R^2$ greater than or equal to 0.994 on all five markets. The CEEMDAN-SE-VMGL and VMD-SE-VMGL pipelines form a second tier, with $R^2$ values ranging from 0.992 to 0.994 and from 0.989 to 0.995, respectively. The EMD-based variants follow, with $R^2$ values between 0.941 and 0.993. The non-decomposition benchmarks LSTM and VMGL achieve consistently low $R^2$ values, between 0.966 and 0.979, and RMSE values 2.5 to 3.2 times those of the proposed model. The EEMD-based models perform worst overall: EEMD-SE-LSTM attains an $R^2$ as low as 0.8477 on CSI~500, RMSE values three to over six times those of the proposed model, and underperforms the non-decomposition benchmarks on all five markets, whereas EEMD-SE-VMGL remains competitive, with $R^2$ between 0.984 and 0.990.

Figure~\ref{FIG:Improvements} shows the percentage improvement of the proposed CEEMDAN-SE-SSA-VMGL model over the eleven benchmark models on RMSE, MAE, and MAPE, quantifying its superiority. The improvement is computed as the relative error reduction achieved by the proposed model over each benchmark on each market. On all markets, CEEMDAN-SE-SSA-LSTM, a model denoised by SSA that shares the same CEEMDAN-SE-SSA front-end as the proposed model, is the closest-performing comparison model within the CEEMDAN family, whereas among the non-decomposition benchmarks, LSTM consistently exhibits the larger performance gap.

On CSI~300, relative to the closest-performing comparison model, the proposed model reduces RMSE, MAE, and MAPE by 8.8\%, 9.0\%, and 8.9\%, respectively, and relative to LSTM by 67.2\%, 67.0\%, and 67.0\%. On CSI~500, the improvements over the closest competitor are 12.4\%, 9.8\%, and 10.1\%, and those over LSTM are 60.8\%, 59.7\%, and 59.9\%. For AAPL, the improvements over the closest competitor are 17.1\%, 16.2\%, and 15.5\%, and those over LSTM are 68.7\%, 67.8\%, and 67.6\%. On GOOGL, the improvements over the closest competitor are 18.1\%, 20.0\%, and 20.9\%, and those over LSTM are 67.7\%, 67.4\%, and 68.0\%. For S\&P~500, the improvements over the closest competitor are 23.7\%, 24.4\%, and 24.2\%, whereas those over LSTM reach as high as 68.5\%, 69.5\%, and 69.5\%. Figure~\ref{FIG:Scatter} shows the scatter plots of the actual versus predicted values, in which the data points cluster tightly around the diagonal on all five datasets without systematic deviation or anomalous dispersion, indicating that the proposed model exhibits stable cross-market predictive consistency.

The results in Table~\ref{tab:model_comparison} and Figure~\ref{FIG:Improvements} confirm that the proposed model has a clear advantage over both simpler standalone architectures and advanced decomposition-based methods. The large performance gap relative to LSTM highlights the limitation of applying deep learning directly to raw financial data in the absence of adequate decomposition and integration optimization. The more modest yet consistent improvement over CEEMDAN-SE-SSA-LSTM demonstrates the additional gain brought by the VMGL predictor, which captures differentiated predictive information across decomposed components more effectively than a single recurrent module. Overall, the proposed framework achieves robust and consistent prediction accuracy under complex, rapidly changing market conditions.

\subsection{Analysis of the role of SSA in mitigating mode mixing}
\begin{figure}
  \centering
   \includegraphics[width=0.80\textwidth]{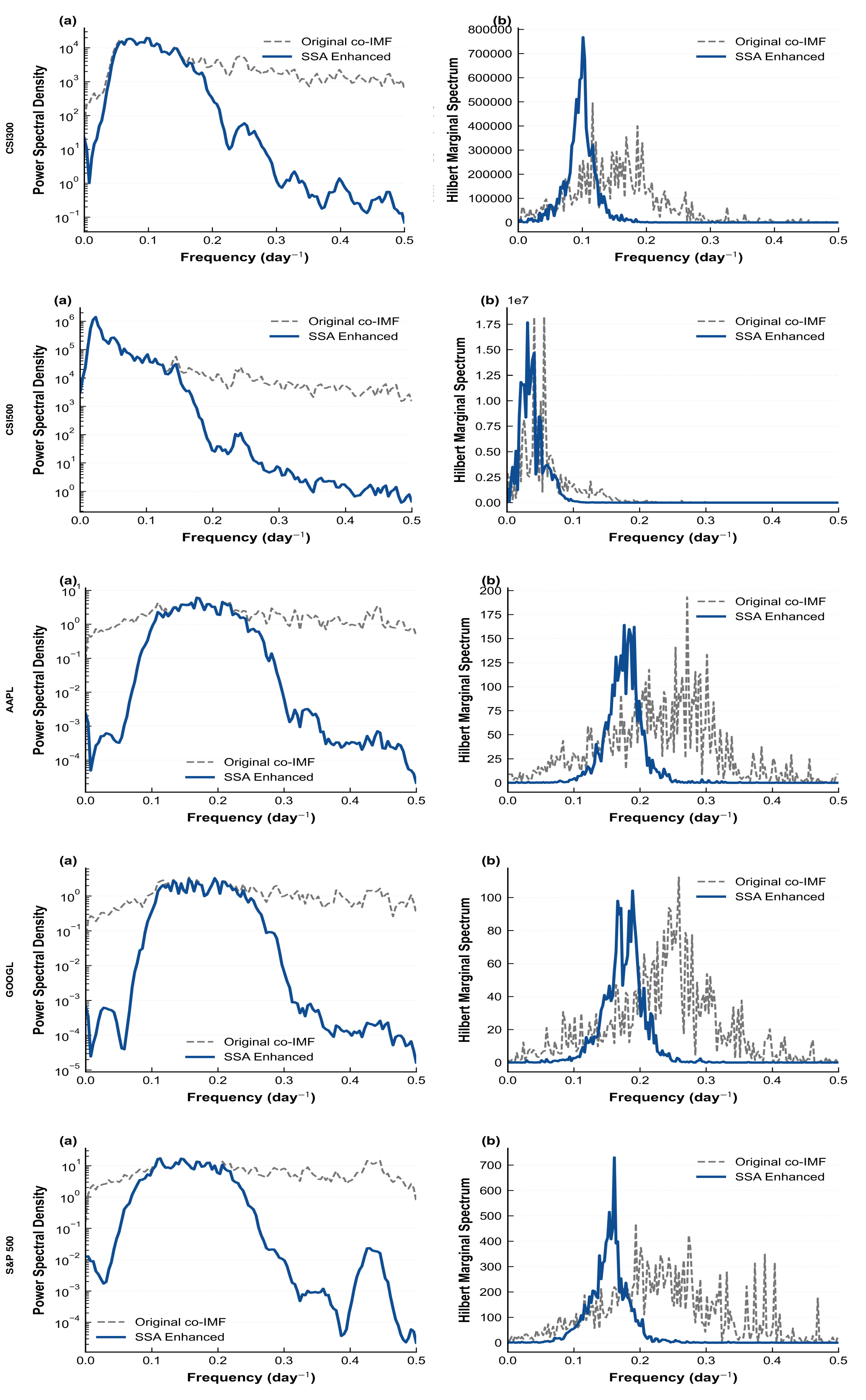}
    \caption{SSA-enhanced processing suppresses mode mixing.}
    \label{FIG:SSA_ENHANCE}
\end{figure}

To further verify the effectiveness of the SSA-enhanced method in suppressing mode mixing and improving signal-decomposition quality, this subsection conducts an in-depth evaluation of the frequency-domain characteristics of the signals. A joint analysis of the power spectral density (PSD) and the Hilbert marginal spectrum reveals significant changes in the frequency-domain characteristics before and after the SSA-enhanced processing. As shown in Figure~\ref{FIG:SSA_ENHANCE}, for the original intrinsic mode function (IMF) series (gray dashed lines), the PSD remains flat over the normalized frequency range $0.0$--$0.5$~day$^{-1}$ on all five datasets. The magnitudes are approximately $10^{1}$ for CSI 300 and CSI 500 and approximately $1$ for AAPL, GOOGL, and S\&P 500, and this nearly uniform distribution indicates a lack of frequency selectivity. The Hilbert marginal spectrum consistently lacks modal-structure features, with the energy distributed over multiple discrete peaks spanning the full frequency range rather than concentrated on a single dominant mode. In CSI 300, the original Hilbert energy exceeds $8 \times 10^{5}$ over multiple wide peaks; in AAPL, GOOGL, and S\&P 500, several spurious peaks reach approximately $175$, approximately $100$, and approximately $400$, respectively. This broadband, multi-peaked feature is consistently observed across all five datasets and is the typical signature of mode mixing: a single IMF captures multiple oscillatory modes at different time scales.

After the SSA-enhanced processing (blue solid lines), both spectra exhibit a consistent transformation across all datasets. In the power spectral density, the previously flat distribution is replaced by sharp, well-defined spectral peaks. For CSI 300, the dominant peak appears at approximately $0.1$~day$^{-1}$ (magnitude approximately $10^{4}$). For AAPL, GOOGL, and S\&P 500, the energy is concentrated within the $0.1$--$0.2$~day$^{-1}$ range (peak magnitude approximately $10^{4}$). CSI 500 exhibits a pronounced low-frequency peak at approximately $0.05$~day$^{-1}$, reaching approximately $10^{7}$. The Hilbert marginal spectrum corroborates this transformation: the original discrete multi-peak pattern is replaced by a smooth unimodal curve. In AAPL and GOOGL, the dominant peak appears at approximately $0.18$~day$^{-1}$ (magnitudes approximately $170$ and approximately $100$, respectively). S\&P 500 exhibits a similar peak at approximately $0.18$~day$^{-1}$ (approximately $650$), whereas CSI 300 shows a peak at approximately $0.1$~day$^{-1}$ (approximately $8 \times 10^{5}$). For CSI 500, the Hilbert energy decays rapidly and approaches zero beyond $0.05$~day$^{-1}$. On all five datasets, the characteristic high-frequency energy in the original spectra (beyond approximately $0.3$~day$^{-1}$) is suppressed to near-zero levels after the SSA processing. The consistency between the PSD and the Hilbert marginal spectrum results indicates that the SSA-enhanced method successfully separates the dominant oscillatory modes from interfering noise and irrelevant frequency components. This markedly reduces mode mixing. The decomposed components thereby acquire clearer physical meaning, more single-scale temporal characteristics, and significantly improved feature purity. These refined components provide a higher-quality input basis for the subsequent training and prediction of the downstream hybrid models, which in turn enhances the convergence stability and long-term prediction accuracy of the models.

\begin{table}
	\centering
	\footnotesize
	\begin{threeparttable}
	\caption{SSA denoising performance comparison across all datasets.}
	\label{Tab:SSA-all}
	\begin{tabular}{c r r r r r r}
		\toprule
		\textbf{Dataset} & 
		\multicolumn{1}{c}{Variance\_Before} & 
		\multicolumn{1}{c}{Variance\_After} & 
		\multicolumn{1}{c}{Variance\_Change(\%)} & 
		\multicolumn{1}{c}{ACF1\_Before} & 
		\multicolumn{1}{c}{ACF1\_After} & 
		\multicolumn{1}{c}{ACF1\_Change} \\
		\midrule
		CSI300 & 968.1809 & 491.2525 & -49.26 & 0.0517 & 0.3543 & 0.3026 \\
		CSI500 & 3343.4916 & 1794.3963 & -46.33 & 0.0846 & 0.3747 & 0.2901 \\
		AAPL & 0.9611 & 0.4482 & -53.37 & 0.1234 & 0.4409 & 0.3174 \\
		GOOGL & 0.6352 & 0.2902 & -54.31 & 0.0867 & 0.4434 & 0.3567 \\
		S\&P500 & 3.3180 & 1.4070 & -57.60 & 0.0968 & 0.4940 & 0.3972 \\
		\bottomrule
	\end{tabular}
	\end{threeparttable}
\end{table}

The PSD and Hilbert marginal spectrum analysis depicted, from the frequency-domain perspective, the marked change in the signal's spectral structure before and after the SSA enhancement. To provide cross-validation in the time domain, Table~\ref{Tab:SSA-all} supplies complementary quantitative evidence of the SSA denoising effect, using the series variance and the first-order autocorrelation coefficient ($\text{ACF}_1$) as the evaluation metrics. Variance characterizes the overall fluctuation amplitude of the series, which effective denoising should reduce markedly; $\text{ACF}_1$ measures the strength of temporal dependence, with structured signals exhibiting strong positive autocorrelation whereas the $\text{ACF}_1$ of white noise approaches zero.

Across the five datasets, the variance reduction lies between $46.33\%$ and $57.60\%$ (mean $52.17\%$), indicating that the SSA processing removes roughly half of the fluctuation energy from each series. At the same time, $\text{ACF}_1$ consistently jumps from a near-zero baseline level before processing ($0.05$--$0.12$) to $0.35$--$0.49$, confirming that the original series is dominated by white noise and that the SSA processing restores a signal with temporal-structure characteristics. The improvements in both indicators exhibit a consistent ordering across the five datasets, namely S\&P 500 $>$ GOOGL $>$ AAPL $>$ CSI 300 $>$ CSI 500, with the mode-mixing mitigation on the US equity markets (variance reduction $53\%$--$58\%$, $\text{ACF}_1$ increase $+0.32$--$0.40$) systematically superior to that on the Chinese equity indices (variance reduction $46\%$--$49\%$, $\text{ACF}_1$ increase $+0.29$--$0.30$). The simultaneity of the variance reduction and the autocorrelation enhancement forms a complementary evidence chain: the former confirms the effective removal of irrelevant volatility components, and the latter verifies that the retained components possess stronger temporal-structure characteristics. These time-domain indicators corroborate the frequency-domain observations above, together demonstrating that the SSA-enhanced method effectively suppresses the noise-dominated unstructured fluctuations in the financial series and successfully separates their dominant oscillatory modes.

\subsection{Analysis of the role of the risk gate in forecast stability}

\begin{figure}
  \centering
   \includegraphics[width=0.95\textwidth]{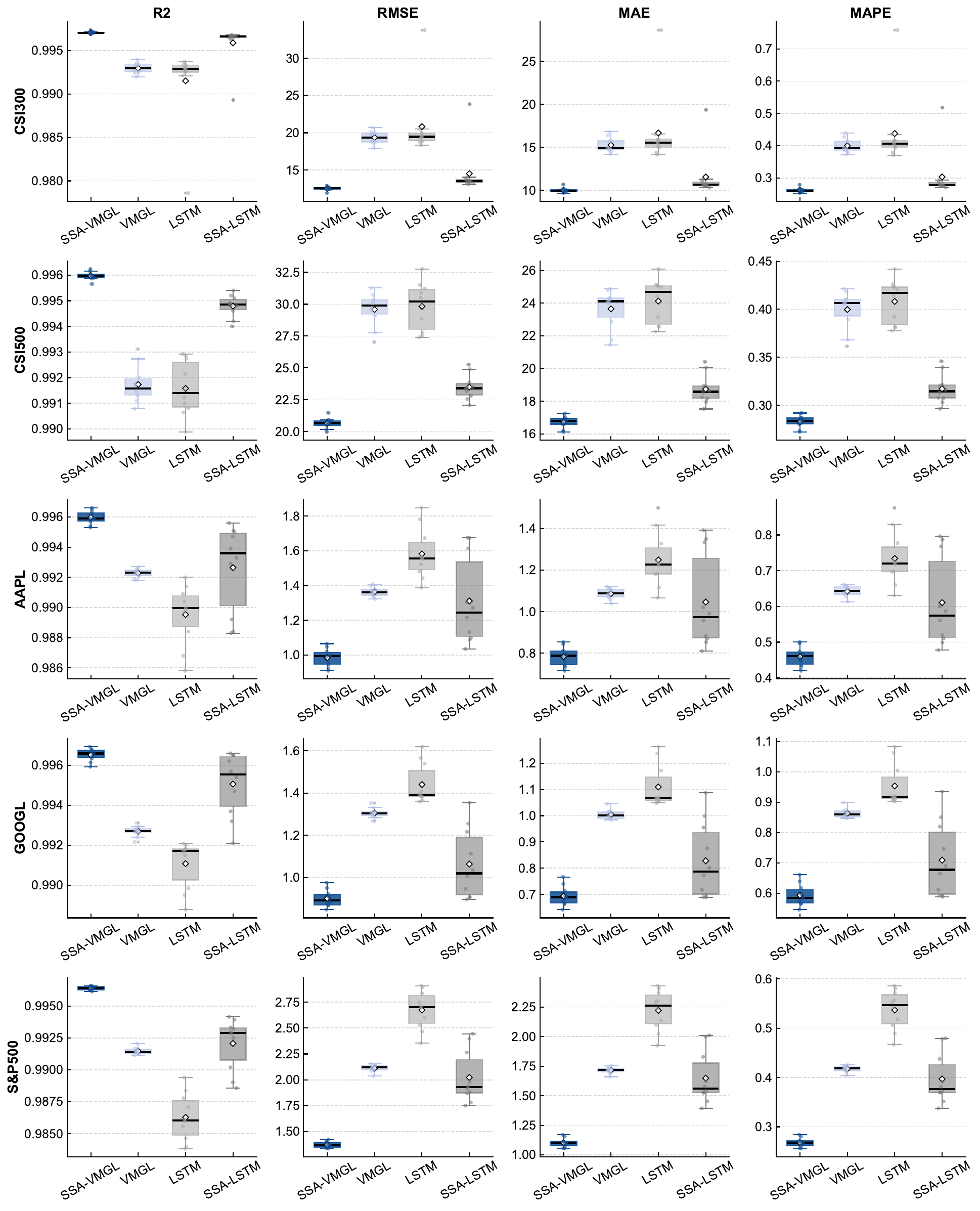}
    \caption{Performance comparison of different models across five datasets in terms of $R^2$, RMSE, MAE, and MAPE.}
    \label{FIG:Boxplot}
\end{figure}

Forecasting accuracy alone cannot guarantee the practical value of a model. Maintaining prediction stability across independent training runs is equally crucial for financial decision-making, because unstable model outputs can undermine portfolio strategies.
Figure~\ref{FIG:Boxplot} shows the distributions of the prediction metrics of four model variants over five datasets, each obtained from 10 independent runs. The interquartile range (IQR) of the full CEEMDAN-SE-SSA-VMGL model is consistently narrower than that of CEEMDAN-SE-SSA-LSTM across all five datasets and all four evaluation metrics, indicating that the VMGL model effectively reduces the run-to-run variability of prediction accuracy.

This stabilization effect is particularly pronounced on the US equity datasets (including both individual stocks and indices). For GOOGL, the interquartile range of MAPE shrinks from approximately \(0.10\) (CEEMDAN-SE-SSA-LSTM) to approximately \(0.04\) (CEEMDAN-SE-SSA-VMGL), with the whisker range narrowing correspondingly. AAPL and S\&P~500 follow a similar pattern, with the SSA-VMGL boxplots containing fewer outliers and a higher concentration of values around the median. On the Chinese equity-index datasets (CSI300, CSI500), because of the larger overall price-scale magnitude, the absolute error ranges (RMSE/MAE) are markedly higher than on the US equity datasets, but the reduction in relative dispersion brought by VMGL remains clearly visible. The comparison between CEEMDAN-SE-VMGL (with SSA removed) and CEEMDAN-SE-LSTM (with both SSA and VMGL removed) shows that the stability gain of VMGL persists even in the absence of SSA denoising.
This reveals that the gating mechanism of the proposed risk gate directly regulates the sensitivity of the recurrent state to input perturbations, rather than merely amplifying the denoising effect of the preprocessing.
The boxplot distributions demonstrate that VMGL not only helps to reduce the average prediction error but also yields more reliable, reproducible prediction results across independent training runs, a property of considerable practical value for prediction systems in real deployment.

\subsection{Statistical significance testing}
To assess whether the accuracy gains of the proposed model are statistically robust, we conducted one-sided Diebold--Mariano tests \citep{Diebold1995} on all five datasets, comparing CEEMDAN-SE-SSA-VMGL pairwise against the eleven benchmark models. Based on the difference in mean squared error loss, we tested the null hypothesis that the proposed model is not superior to the comparison model, against the one-sided alternative. The results are summarized in Table~\ref{tab:DM_test}. Across all $5 \times 11 = 55$ comparisons, the proposed model achieves a significantly lower mean squared error at the 1\% significance level in every case. All Diebold--Mariano statistics are negative, reflecting that the proposed model produces smaller squared prediction errors than each benchmark on all datasets. Even the most competitive comparison models, CEEMDAN-SE-SSA-LSTM and VMD-SE-VMGL, have $p$-values below 0.01 on all datasets, confirming that the performance gap is significant not only against the weaker baselines but also when facing the most competitive alternatives.

\begin{table}[htbp]
	\centering
	\footnotesize
	\setlength{\tabcolsep}{1.5pt}
	\begin{threeparttable}
		\caption{Diebold--Mariano test results of the proposed CEEMDAN-SE-SSA-VMGL model versus other benchmark models.}
		\label{tab:DM_test}
		\begin{tabular}{l cc cc cc cc cc}
			\toprule
			\multirow{2}{*}{Model} & \multicolumn{2}{c}{S\&P500} & \multicolumn{2}{c}{AAPL} & \multicolumn{2}{c}{CSI300} & \multicolumn{2}{c}{CSI500} & \multicolumn{2}{c}{GOOGL} \\
			\cmidrule(lr){2-11}
			& Stat. & p-value & Stat. & p-value & Stat. & p-value & Stat. & p-value & Stat. & p-value \\
			\midrule
			LSTM & -10.67*** & 0.0000 & -9.36*** & 0.0000 & -10.33*** & 0.0000 & -8.91*** & 0.0000 & -6.92*** & 0.0000 \\
			VMGL & -11.70*** & 0.0000 & -8.88*** & 0.0000 & -9.46*** & 0.0000 & -8.96*** & 0.0000 & -6.06*** & 0.0000 \\
			EMD-SE-LSTM & -7.74*** & 0.0000 & -10.26*** & 0.0000 & -17.73*** & 0.0000 & -6.81*** & 0.0000 & -9.19*** & 0.0000 \\
			EMD-SE-VMGL & -4.49*** & 0.0000 & -9.92*** & 0.0000 & -9.99*** & 0.0000 & -5.83*** & 0.0000 & -7.77*** & 0.0000 \\
			EEMD-SE-LSTM & -11.33*** & 0.0000 & -11.64*** & 0.0000 & -9.31*** & 0.0000 & -9.95*** & 0.0000 & -10.22*** & 0.0000 \\
			EEMD-SE-VMGL & -9.25*** & 0.0000 & -9.03*** & 0.0000 & -9.72*** & 0.0000 & -8.22*** & 0.0000 & -7.45*** & 0.0000 \\
			VMD-SE-LSTM & -11.85*** & 0.0000 & -8.58*** & 0.0000 & -11.72*** & 0.0000 & -7.92*** & 0.0000 & -8.28*** & 0.0000 \\
			VMD-SE-VMGL & -8.43*** & 0.0000 & -3.07*** & 0.0011 & -6.81*** & 0.0000 & -7.13*** & 0.0000 & -7.16*** & 0.0000 \\
			CEEMDAN-SE-LSTM & -9.11*** & 0.0000 & -7.17*** & 0.0000 & -7.52*** & 0.0000 & -6.55*** & 0.0000 & -7.02*** & 0.0000 \\
			CEEMDAN-SE-VMGL & -7.16*** & 0.0000 & -6.95*** & 0.0000 & -7.23*** & 0.0000 & -5.04*** & 0.0000 & -6.15*** & 0.0000 \\
			CEEMDAN-SE-SSA-LSTM & -6.15*** & 0.0000 & -5.56*** & 0.0000 & -2.86*** & 0.0021 & -2.69*** & 0.0036 & -3.55*** & 0.0002 \\
			\bottomrule
		\end{tabular}
		\begin{tablenotes}[flushleft]
			\small
			\item \textbf{Note:} *, **, and *** respectively represent rejection of the null hypothesis at the 10\%, 5\%, and 1\% significance levels. $H_0$: The CEEMDAN-SE-SSA-VMGL model proposed in our study does not have better predictive ability than the comparison model. $H_1$: The CEEMDAN-SE-SSA-VMGL model proposed in our study has significantly better predictive ability than the comparison model. Statistics are reported to two decimal places and p-values to four decimal places.
		\end{tablenotes}
	\end{threeparttable}
\end{table}

As a complement to the pairwise Diebold--Mariano tests, we further employ the model-confidence-set (MCS) procedure proposed by \citet{Hansen2011} to evaluate the relative predictive performance of all models. This method identifies, without requiring a pre-specified benchmark, a subset of models that contains the best-performing model at a given confidence level. At each step, two complementary statistics are used to test the null hypothesis that all remaining models in the set have equal predictive ability. The range statistic $T_R$ captures the largest single pairwise difference, whereas the semi-quadratic statistic $T_{SQ}$ aggregates the squared deviations across all model pairs. Models are eliminated sequentially until the remaining set cannot be rejected at the 10\% significance level. The model-confidence-set results are reported in Table~\ref{tab:MCS_test}. On all five datasets, the proposed CEEMDAN-SE-SSA-VMGL model is the only model retained in the confidence set, with $p$-values of 1.0000 under both the $T_R$ and $T_{SQ}$ statistics. All other models, including the best-performing CEEMDAN-family alternatives CEEMDAN-SE-VMGL, CEEMDAN-SE-LSTM, and CEEMDAN-SE-SSA-LSTM, are eliminated with $p$-values far below the 0.10 threshold. The elimination outcomes under $T_R$ and $T_{SQ}$ remain highly consistent across all datasets, further confirming the robustness of the elimination process: no model is retained under one statistic yet rejected under the other. The mutual corroboration of the pairwise Diebold--Mariano tests and the model-confidence-set tests demonstrates that the proposed CEEMDAN-SE-SSA-VMGL model achieves statistically robust and consistent superiority over all benchmark models on all five datasets.
\begin{table}[htbp]
	\centering
	\footnotesize
	\setlength{\tabcolsep}{1.2pt}
	\begin{threeparttable}
		\caption{Model Confidence Set (MCS) test results.}
		\label{tab:MCS_test}
		\begin{tabular}{l cc cc cc cc cc}
			\toprule
			\multirow{2}{*}{Model} & \multicolumn{2}{c}{S\&P500} & \multicolumn{2}{c}{AAPL} & \multicolumn{2}{c}{CSI300} & \multicolumn{2}{c}{CSI500} & \multicolumn{2}{c}{GOOGL} \\
			\cmidrule(lr){2-11}
			& $T_R$ & $T_{SQ}$ & $T_R$ & $T_{SQ}$ & $T_R$ & $T_{SQ}$ & $T_R$ & $T_{SQ}$ & $T_R$ & $T_{SQ}$ \\
			\midrule
			LSTM & 0.0000 & 0.0000 & 0.0000 & 0.0000 & 0.0000 & 0.0000 & 0.0000 & 0.0000 & 0.0000 & 0.0000 \\
			VMGL & 0.0000 & 0.0000 & 0.0000 & 0.0000 & 0.0000 & 0.0000 & 0.0000 & 0.0000 & 0.0000 & 0.0000 \\
			EMD-SE-LSTM & 0.0000 & 0.0000 & 0.0000 & 0.0000 & 0.0000 & 0.0000 & 0.0000 & 0.0000 & 0.0000 & 0.0000 \\
			EMD-SE-VMGL & 0.0010 & 0.0010 & 0.0000 & 0.0000 & 0.0000 & 0.0000 & 0.0000 & 0.0000 & 0.0000 & 0.0000 \\
			EEMD-SE-LSTM & 0.0000 & 0.0000 & 0.0000 & 0.0000 & 0.0000 & 0.0000 & 0.0000 & 0.0000 & 0.0000 & 0.0000 \\
			EEMD-SE-VMGL & 0.0000 & 0.0000 & 0.0000 & 0.0000 & 0.0000 & 0.0000 & 0.0000 & 0.0000 & 0.0000 & 0.0000 \\
			VMD-SE-LSTM & 0.0005 & 0.0000 & 0.0005 & 0.0000 & 0.0000 & 0.0000 & 0.0000 & 0.0000 & 0.0000 & 0.0000 \\
			VMD-SE-VMGL & 0.0000 & 0.0000 & 0.0055 & 0.0105 & 0.0000 & 0.0000 & 0.0000 & 0.0000 & 0.0000 & 0.0000 \\
			CEEMDAN-SE-LSTM & 0.0000 & 0.0000 & 0.0000 & 0.0000 & 0.0000 & 0.0000 & 0.0000 & 0.0000 & 0.0000 & 0.0000 \\
			CEEMDAN-SE-VMGL & 0.0000 & 0.0000 & 0.0000 & 0.0000 & 0.0000 & 0.0000 & 0.0000 & 0.0000 & 0.0000 & 0.0000 \\
			CEEMDAN-SE-SSA-LSTM & 0.0000 & 0.0000 & 0.0010 & 0.0005 & 0.0145 & 0.0150 & 0.0145 & 0.0215 & 0.0000 & 0.0010 \\
			CEEMDAN-SE-SSA-VMGL & \textbf{\underline{1.0000}} & \textbf{\underline{1.0000}} & \textbf{\underline{1.0000}} & \textbf{\underline{1.0000}} & \textbf{\underline{1.0000}} & \textbf{\underline{1.0000}} & \textbf{\underline{1.0000}} & \textbf{\underline{1.0000}} & \textbf{\underline{1.0000}} & \textbf{\underline{1.0000}} \\
			\bottomrule
		\end{tabular}
		\begin{tablenotes}[flushleft]
			\small
			\item \textbf{Note:} Numbers in bold denote the model with the best predictive performance under the MCS criterion. Numbers with a $p$-value greater than 0.1 are underlined and retained in the MCS. The null hypothesis assumes that all models under consideration exhibit equivalent predictive performance. Rejection of the null implies that at least one model demonstrates predictive performance that is statistically distinct from the others.
		\end{tablenotes}
	\end{threeparttable}
\end{table}

\subsection{Alternative windows}
\begin{table}[htbp]
  \centering
  \begin{threeparttable}
  \caption{Best-run metrics per dataset across sliding window sizes $H$.}
  \label{tab:sensitivity_pivot}
  \begin{tabular}{l l c c c c}
    \toprule
    Dataset & Metric & $H=5$ & $H=10$ & $H=15$ & $H=20$ \\
    \midrule
    \multirow{4}{*}{CSI 300} & $R^2$ & 0.996932 & \textbf{0.997338} & 0.997163 & 0.997034 \\
     & RMSE & 12.7805 & \textbf{11.9096} & 12.1404 & 12.2485 \\
     & MAE & 10.3809 & \textbf{9.6850} & 9.9169 & 9.9396 \\
     & MAPE (\%) & 0.2714 & \textbf{0.2534} & 0.2591 & 0.2607 \\
    \cmidrule{1-6}
    \multirow{4}{*}{CSI 500} & $R^2$ & 0.996156 & \textbf{0.996237} & 0.995996 & 0.995863 \\
     & RMSE & 20.0087 & \textbf{19.9813} & 20.6772 & 20.8890 \\
     & MAE & 16.2935 & \textbf{16.1913} & 16.8067 & 16.6059 \\
     & MAPE (\%) & 0.2759 & \textbf{0.2732} & 0.2843 & 0.2818 \\
    \cmidrule{1-6}
    \multirow{4}{*}{AAPL} & $R^2$ & 0.996191 & \textbf{0.996705} & 0.995843 & 0.995323 \\
     & RMSE & 1.0137 & \textbf{0.8908} & 0.9496 & 0.9652 \\
     & MAE & 0.8099 & \textbf{0.7180} & 0.7513 & 0.7685 \\
     & MAPE (\%) & 0.4739 & \textbf{0.4233} & 0.4401 & 0.4465 \\
    \cmidrule{1-6}
    \multirow{4}{*}{GOOGL} & $R^2$ & \textbf{0.996956} & 0.996924 & 0.996691 & 0.996754 \\
     & RMSE & 0.8647 & 0.8472 & 0.8638 & \textbf{0.8425} \\
     & MAE & 0.6612 & \textbf{0.6415} & 0.6658 & 0.6550 \\
     & MAPE (\%) & 0.5658 & \textbf{0.5453} & 0.5668 & 0.5549 \\
    \cmidrule{1-6}
    \multirow{4}{*}{S\&P 500} & $R^2$ & 0.994957 & \textbf{0.996590} & 0.996333 & 0.996079 \\
     & RMSE & 1.6451 & \textbf{1.3356} & 1.3635 & 1.4008 \\
     & MAE & 1.3478 & \textbf{1.0528} & 1.0973 & 1.1258 \\
     & MAPE (\%) & 0.3263 & \textbf{0.2556} & 0.2664 & 0.2719 \\
    \bottomrule
  \end{tabular}
  \begin{tablenotes}[flushleft]
  \footnotesize
  \item \textbf{Note:} Bold values denote the best performance for each metric across the four window sizes. For $R^2$, a higher value is better; for RMSE, MAE, and MAPE, a lower value is better.
  \end{tablenotes}
  \end{threeparttable}
\end{table}

The sliding-window length determines the historical-information capacity of the input sequence and is a key hyperparameter affecting model prediction performance. To assess the robustness of the proposed framework to window size, this section systematically compares four sliding-window configurations ($H = 5$, $10$, $15$, and $20$), with the results summarized in Table~\ref{tab:sensitivity_pivot}.

Overall, the prediction performance of the model does not fluctuate greatly with drastic changes in window size. Across all five datasets, the $R^2$ values under the four window configurations remain stably above 0.994, with the largest span being only 0.0016 on S\&P 500 (the difference between 0.9966 at $H = 10$ and 0.9950 at $H = 5$). Similarly, the maximum fluctuation of MAPE across the four windows is only 0.011 percentage points on CSI 500 and 0.071 percentage points on S\&P 500. Taking S\&P 500 as an example, $R^2$ stays within the narrow range of 0.9950 to 0.9966 from $H = 5$ to $H = 20$, RMSE ranges between 1.34 and 1.65, and MAPE ranges between 0.26\% and 0.33\%. This high consistency across window configurations indicates that the proposed framework does not exhibit sensitive dependence on the sliding-window hyperparameter, demonstrating sound robustness.

Although the optimal window size differs slightly across datasets (four datasets attain the optimum at $H = 10$, whereas GOOGL attains its optimal metrics at $H = 5$ for $R^2$, $H = 10$ for MAE and MAPE, and $H = 20$ for RMSE), the performance degradation between the best and worst results is extremely limited, further corroborating the reliability of the model under a limited hyperparameter search. These results demonstrate that the intrinsic stability conferred on the CEEMDAN-SE-SSA-VMGL framework by decomposition--reconstruction and volatility-adaptive modulation enables competitive prediction accuracy in real deployment without per-market fine-tuning.

\subsection{Computational time}

\begin{table}[h]
\centering
\small
\begin{threeparttable}
\caption{Total running time in seconds.}
\label{tab:running_time}
\begin{tabular}{lrrrrr}
\toprule
\textbf{Model} & \textbf{CSI 300} & \textbf{CSI 500} & \textbf{AAPL} & \textbf{GOOGL} & \textbf{S\&P 500} \\
\midrule
LSTM & 100.4 & 98.4 & 27.6 & 75.7 & 104.0 \\
VMGL & 72.1 & 91.0 & 64.2 & 77.6 & 95.5 \\
EMD-SE-LSTM & 599.9 & 658.4 & 686.5 & 653.2 & 597.6 \\
EMD-SE-VMGL & 763.9 & 831.4 & 752.5 & 801.2 & 775.8 \\
EEMD-SE-LSTM & 699.4 & 732.5 & 675.1 & 634.0 & 682.0 \\
EEMD-SE-VMGL & 852.4 & 898.2 & 767.5 & 793.8 & 768.5 \\
VMD-SE-LSTM & 783.1 & 786.5 & 829.3 & 872.5 & 833.6 \\
VMD-SE-VMGL & 893.0 & 953.6 & 816.8 & 831.4 & 955.8 \\
CEEMDAN-SE-LSTM & 708.7 & 783.5 & 750.5 & 696.9 & 718.3 \\
CEEMDAN-SE-VMGL & 894.6 & 934.2 & 859.6 & 854.2 & 967.9 \\
CEEMDAN-SE-SSA-LSTM & 2532.8 & 2238.0 & 1376.0 & 1590.5 & 2083.5 \\
CEEMDAN-SE-SSA-VMGL & 1039.7 & 1015.2 & 1032.9 & 1010.6 & 1429.9 \\
\bottomrule
\end{tabular}
\end{threeparttable}
\end{table}
	
To examine the deployment feasibility of the model in real financial scenarios, this section further evaluates the computational cost of each model on the basis of the validated prediction performance. All models are trained for 200 epochs with the same early-stopping patience, and the unified sample-entropy clustering ensures that all sample-entropy-based models generate the same number of components. These controls attribute the differences in running time strictly to the choice of model architecture, rather than to differences in the training budget or the number of components, ensuring a fair and reproducible comparison.

The running times of the models are reported in Table~\ref{tab:running_time}. The standalone architectures exhibit the shortest running times. The standard LSTM and VMGL predictors take about 81 s on average per dataset. Such models process the raw series directly through a single recurrent layer and converge quickly without signal decomposition. The running time of the single-decomposition models increases substantially. The EMD and EEMD variants are the fastest, with their LSTM and VMGL variants averaging between 639 and 816 s, followed by VMD at 821--890 s and CEEMDAN at 732--902 s. Relative to single-trial EMD, the CEEMDAN ensemble introduces about 15\% additional computational overhead. Among all decomposition strategies, the VMGL architecture, which introduces the extra risk gate, incurs an 8--23\% per-epoch computational overhead relative to the standard LSTM.

Introducing SSA denoising further prolongs the CEEMDAN-pipeline processing time. The CEEMDAN-SE-SSA-LSTM model trains a standard LSTM on each component, taking on average 1964 s. In contrast, the proposed CEEMDAN-SE-SSA-VMGL model averages only 1106 s, running about 1.78 times as fast as its LSTM counterpart, despite sharing exactly the same CEEMDAN decomposition, sample-entropy clustering, and SSA modules. This result reverses the pattern observed before SSA was introduced, namely that without the SSA stage, the VMGL variant is always 8--23\% slower than its LSTM counterpart. Introducing SSA increases the running time of VMGL by only 204 s, i.e., 23\%, whereas the LSTM variant increases by 1233 s, i.e., 168\%. The proposed model therefore preserves the accuracy gain of SSA denoising at a small fraction of the time cost incurred by the standard LSTM design.

\subsection{Backtesting and trade simulation}

\begin{table}[tb]
	\centering
	\small
	\begin{threeparttable}
		\caption{Sharpe ratio performance by investment strategy.}
		\label{TAB:sharpe_ratios}
		\begin{tabular}{lccccc}
			\toprule
			Strategy & CSI300 & CSI500 & AAPL & GOOGL & S\&P500 \\
			\midrule
			\multicolumn{6}{l}{\textit{Buy-and-Hold}} \\
			\quad Buy-and-Hold & $-$0.0959 & $-$0.0633 & 0.1205 & 0.0924 & 0.1083 \\
			\midrule
			\multicolumn{6}{l}{\textit{Prediction-driven directional strategy}} \\
			CEEMDAN-SE-SSA-VMGL & {\bf 0.5357} & {\bf 0.6065} & {\bf 0.6630} & {\bf 0.6214} & {\bf 0.6612} \\
			CEEMDAN-SE-SSA-LSTM & \underline{0.5153} & \underline{0.5783} & \underline{0.6115} & \underline{0.5931} & \underline{0.6193} \\
			VMD-SE-VMGL & 0.4411 & 0.4228 & 0.5986 & 0.5366 & 0.5035 \\
			CEEMDAN-SE-VMGL & 0.4270 & 0.4891 & 0.5425 & 0.5382 & 0.5433 \\
			VMD-SE-LSTM & 0.0924 & 0.3597 & 0.5160 & 0.4954 & 0.4477 \\
			CEEMDAN-SE-LSTM & 0.4340 & 0.4658 & 0.5440 & 0.5342 & 0.4870 \\
			EMD-SE-VMGL & 0.2413 & 0.5135 & 0.4035 & 0.3238 & 0.5746 \\
			EMD-SE-LSTM & $-$0.0516 & 0.4916 & 0.3625 & 0.2552 & 0.5801 \\
			EEMD-SE-VMGL & 0.3217 & 0.3956 & 0.4865 & 0.4568 & 0.4373 \\
			EEMD-SE-LSTM & 0.1138 & 0.1299 & 0.2850 & 0.3619 & 0.2779 \\
			LSTM & $-$0.0870 & $-$0.0271 & $-$0.0186 & 0.0220 & 0.1042 \\
			VMGL & $-$0.1082 & $-$0.0871 & 0.0407 & 0.0237 & 0.0090 \\
			\bottomrule
		\end{tabular}
		\begin{tablenotes}[flushleft]
			\footnotesize
			\item \textbf{Note:} Bold and underlined values respectively denote the highest and second-highest Sharpe ratio in each column. 
		\end{tablenotes}
	\end{threeparttable}
\end{table}

\begin{figure}
  \centering
   \includegraphics[width=0.95\textwidth]{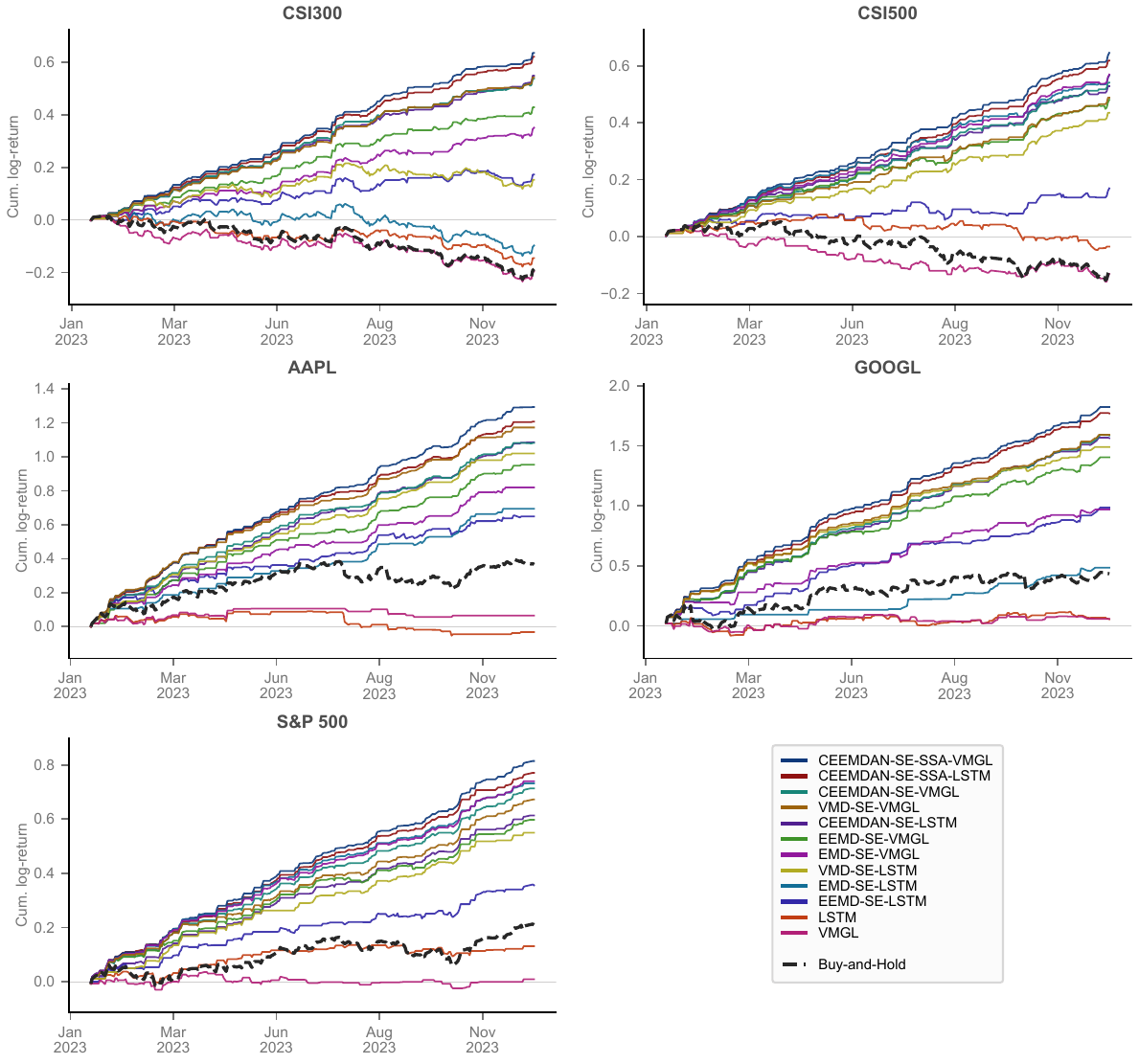}
    \caption{Cumulative log-returns of the signal-based long/flat strategy across five datasets.}
    
    \label{FIG:Returns}
\end{figure}

To evaluate the real-world utility of the forecasting models beyond traditional point-prediction accuracy, we assess their economic value through an explicit investment strategy. Excellent statistical performance on error-based metrics does not guarantee profitable trading outcomes, and profitability analysis provides concrete evidence of a model's reliability in real-world financial decision-making. We implement a prediction-driven directional strategy: on each trading day \(t\), if the model's prediction \(\hat{P}_t\) exceeds the previous day's closing price \(P_{t-1}\), a full long position is established; otherwise, the position is kept flat. Under the assumptions of zero transaction costs, zero slippage, and a zero risk-free rate, the daily log return is computed as the product of \(\ln(P_t / P_{t-1})\) and the position weight. The passive buy-and-hold strategy serves as the benchmark.

Table~\ref{TAB:sharpe_ratios} reports the daily Sharpe ratios of each model on the five benchmark datasets. This metric measures risk-adjusted return, with higher values indicating superior investment performance for equal risk. A positive Sharpe ratio implies that the return exceeds the risk-free rate, whereas a negative value indicates that the strategy underperformed the risk-free asset over the evaluation window. When driven by the decomposition-enhanced forecasting models, this prediction-driven directional strategy consistently generates risk-adjusted returns superior to those of the passive buy-and-hold benchmark across all five markets. The buy-and-hold benchmark yields Sharpe ratios of \(-0.0959\) and \(-0.0633\) on CSI300 and CSI500, respectively, reflecting the adverse market environment faced by the Chinese equity market within the 2023 test window, whereas it produces mildly positive values of \(0.1205\), \(0.0924\), and \(0.1083\) on AAPL, GOOGL, and S\&P~500, respectively. In contrast, all CEEMDAN-based models convert these passive results into strong positive active returns, with Sharpe ratios ranging from 0.4270 to 0.6630 across the benchmarks, fully demonstrating the practical value of signal-decomposition-enhanced forecasting models in guiding investment-timing decisions.

Among all evaluated methods, the CEEMDAN-SE-SSA-VMGL model attains the highest Sharpe ratio on all five benchmarks, with values of \(0.5357\), \(0.6065\), \(0.6630\), \(0.6214\), and \(0.6612\), respectively, whereas CEEMDAN-SE-SSA-LSTM ranks second on every benchmark with \(0.5153\), \(0.5783\), \(0.6115\), \(0.5931\), and \(0.6193\). A clear hierarchy emerges among the different signal-decomposition families, with the CEEMDAN-based models achieving the highest average Sharpe ratio across datasets, followed by the VMD-based, EMD-based, and EEMD-based models. The pure deep-learning benchmark models LSTM and VMGL, which run directly on the raw price series without any signal decomposition, achieve average Sharpe ratios of approximately \(-0.001\) and \(-0.024\), respectively, and neither beats buy-and-hold on the remaining benchmarks (except that LSTM slightly outperforms on the Chinese markets, where the buy-and-hold benchmark is negative). This performance gradient confirms signal decomposition as the decisive factor in achieving profitable directional trading, and also indicates that the CEEMDAN-SE-SSA-VMGL pipeline provides the most effective decomposition--prediction combination across diverse market conditions.

Figure~\ref{FIG:Returns} visualizes the cumulative log returns of all strategies over the test period. The CEEMDAN-based strategies consistently trace the steepest upward trajectories in every subfigure, whereas the pure LSTM and VMGL strategies stay at or below the zero line throughout. The buy-and-hold reference curves decline on the Chinese equity indices and rise only mildly on the US equities, remaining far below all decomposition-enhanced strategies throughout. The within-family performance ordering is largely preserved across datasets, indicating that the decomposition--prediction pipeline captures generalizable directional signals rather than overfitting any single asset. The absolute Sharpe ratios reported here represent upper-bound estimates under frictionless-execution assumptions; however, the relative ordering among the models, especially the consistent dominance exhibited by the CEEMDAN-based architectures, is robust to these simplifying assumptions, because all strategies are evaluated under exactly identical conditions.

In summary, the profitability analysis demonstrates that the proposed CEEMDAN-SE-SSA-VMGL model, which integrates decomposition, high-frequency reconstruction, and a volatility-adaptive mechanism, translates superior point-prediction accuracy into economically meaningful trading gains on all five benchmark markets. The consistent results presented by the Sharpe ratios and cumulative-return trajectories indicate that SSA-based denoising and the volatility-adaptive gating mechanism are the core elements endowing the proposed model with practical investment value, whereas the CEEMDAN signal decomposition constitutes the key preprocessing backbone supporting the framework.
\section{Conclusion}
\label{sec5}
This paper proposes the CEEMDAN-SE-SSA-VMGL hybrid framework to address the challenges of multi-scale, high-noise, and non-stationary volatility in financial time-series forecasting. The framework integrates adaptive signal decomposition, entropy-guided component clustering, singular-spectrum-analysis-based high-frequency denoising, and a volatility-modulated GARCH-LSTM network. Its core innovations are twofold. First, applying singular spectrum analysis to the high-frequency intrinsic mode functions of the financial series alleviates mode mixing and suppresses the residual noise that standard CEEMDAN fails to eliminate. Second, embedding a differentiable GARCH layer inside the long short-term memory architecture as a risk gate endows the recurrent unit with intrinsic robustness to non-stationary volatility without sacrificing end-to-end trainability. These two designs jointly advance the hybrid financial-forecasting paradigm by addressing the data-side denoising and model-side volatility-adaptation problems that previous decomposition-based frameworks have tended to handle in isolation.

Empirical evaluation across five representative financial series from the US and Chinese markets, namely CSI 300, CSI 500, AAPL, GOOGL, and S\&P 500, shows that CEEMDAN-SE-SSA-VMGL consistently outperforms eleven benchmark models on all error metrics. Pairwise Diebold--Mariano tests and the model-confidence-set procedure together confirm the statistical significance of this advantage. Spectral and time-domain analyses confirm that the singular-spectrum-analysis post-processing of the high-frequency components suppresses mode mixing and restores structured time-series signals from noise-dominated modes. The VMGL risk gate reduces the dispersion of prediction accuracy across multiple independent training runs and achieves a better accuracy-to-computational-cost ratio than the singular-spectrum-analysis-enhanced standard LSTM pipeline. Beyond point-prediction accuracy, the directional trading strategy driven by the model predictions attains Sharpe ratios that rank first on all five datasets. Its realized returns substantially exceed those of the same strategy driven by any competing model and those of the buy-and-hold benchmark. These results demonstrate the model's practical value in portfolio construction and risk management. Taken together, these findings indicate that combining rigorous signal denoising with volatility-aware sequence modeling can significantly enhance the accuracy, stability, and economic value of financial time-series forecasting.

Several limitations of this study should be acknowledged. First, the model relies solely on univariate closing-price input features and has not yet been extended to other asset classes such as cryptocurrencies, commodities, or foreign-exchange markets; evaluation across asset classes would help verify the broader generality of the method. Second, although the framework performs well over the whole test period, its prediction robustness during periods of extreme market impact, such as financial crises or sudden policy events, remains to be systematically examined. Finally, this study uses only historical price data and does not incorporate external features such as macroeconomic indicators, investor-sentiment indices, or news-text signals; future work could introduce these features to enrich the model's information input and further improve prediction accuracy.

Looking ahead, several promising directions warrant further exploration. Extending the evaluation to intraday and high-frequency trading scenarios would test the framework's responsiveness under rapid market-regime switching. Introducing alternative volatility estimators, such as realized volatility or option-implied volatility, into the risk gate may further optimize the model's adaptation to extreme events. Exploring the transfer of the singular-spectrum-analysis-denoised decomposition pipeline to other asset classes and to multi-asset portfolio forecasting would help broaden its applicability. Generalizing the activation-level end-to-end volatility-integration scheme to a wider range of deep-learning architectures, and examining its adaptability across diverse financial-forecasting scenarios, would verify the generality and transferability of the integration paradigm and lay the foundation for a cross-paradigm, cross-market volatility-aware forecasting system. In summary, the CEEMDAN-SE-SSA-VMGL framework represents a solid step toward building hybrid models capable of withstanding volatility and promises strong support for both academic research and practical quantitative investment decision-making.

\section*{Data availability}
All data supporting the findings of this study are publicly available and can be retrieved with the Python \texttt{efinance} package (https://github.com/Micro-sheep/efinance).

\section*{Declaration of Competing Interest}
The authors declare that they have no known competing financial interests or personal relationships that could have appeared to influence the work reported in this paper.




\clearpage 

\printcredits





\end{document}